\documentclass[sigconf, nonacm]{acmart}
\usepackage[utf8]{inputenc}
\usepackage{multicol}
\usepackage[inline]{enumitem}
\usepackage{mathtools}
\usepackage{enumitem}
\usepackage{footnote}
\usepackage{xcolor}
\usepackage[bottom]{footmisc}
\usepackage{ltablex,booktabs}
\usepackage[flushleft]{threeparttable}
\usepackage{cuted}
\usepackage{graphicx}
\usepackage[english]{babel} 
\usepackage{blindtext}
\usepackage{makecell}
\usepackage{placeins}

\DeclarePairedDelimiter\ceil{\lceil}{\rceil} 
\DeclarePairedDelimiter\floor{\lfloor}{\rfloor} 
\makesavenoteenv{tabular}
\makesavenoteenv{table}
\begin{document}
\title{Design Trade-offs for a Robust Dynamic Hybrid Hash Join (Extended Version)}

\author{Shiva Jahangiri}
\affiliation{%
  \institution{University of California, Irvine}
}
\email{shivaj@uci.edu}

\author{Michael J. Carey}
\affiliation{%
  \institution{University of California, Irvine}
}
\email{mjcarey@ics.uci.edu}

\author{Johann-Christoph Freytag}
\affiliation{%
  \institution{Humboldt-Universität zu Berlin}
}
\email{freytag@informatik.hu-berlin.de}

\begin{abstract}
\nocite{*}
The Join operator, as one of the most expensive and commonly used operators in database systems, plays a substantial role in Database Management System (DBMS) performance. Among the many different Join algorithms studied over the last decades, Hybrid Hash Join (HHJ) has proven to be one of the most efficient and widely-used join algorithms. While HHJ’s performance depends largely on accurate statistics and information about the input relations, it may not always be practical or possible for a system to have such information available.

HHJ’s design depends on many details to perform well. This paper is an experimental and analytical study of the trade-offs in designing a robust and dynamic HHJ operator. We revisit the design and optimization techniques suggested by previous studies through extensive experiments, comparing them with other algorithms designed by us or used in related studies.

We explore the impact of the number of partitions on HHJ’s performance and propose a lower bound and a default value for the number of partitions. We continue by designing and evaluating different partition insertion techniques to maximize memory utilization with the least CPU cost. In addition, we consider a comprehensive set of algorithms for dynamically selecting a partition to spill and compare the results against previously published studies. We then present two alternative growth policies for spilled partitions and study their effectiveness using experimental and model-based analyses.

These algorithms have been implemented in the context of Apache AsterixDB and evaluated under different scenarios such as variable record sizes, different distributions of join attributes, and different storage types, including HDD, SSD, and Amazon Elastic Block Store (Amazon EBS) \cite{EBS}.

\end{abstract}

\maketitle

\section{Introduction}

As one of the most popular and expensive DBMS operators, the join operator can significantly impact the performance of a DBMS. HHJ \cite{DBLP:journals/tods/Shapiro86} has shown superior performance in computing the equijoin of two datasets among other kinds of join operators. In a nutshell, HHJ groups the records of each dataset into disjoint partitions. A hash table is created to hold one of the partitions in memory (memory-resident partition), while the rest will be written (spilled) to disk to be processed in the next rounds of HHJ. The number of the partitions and the selection of the memory-resident partition are static decisions made at the compile time of an HHJ operator. While previous studies \cite{DBLP:journals/tods/Shapiro86,DBLP:journals/vldb/HaasCLS97} have suggested various cost models and optimization techniques for enhancing such decisions, these studies have two shortcomings: \begin{enumerate*}
    \item They assume a uniform distribution for join attribute values.
    \item Their cost models rely on having accurate statistical information such as input sizes prior to query execution.
\end{enumerate*}

Unfortunately, collecting and accessing or predicting such information may not always be feasible. For example: 

\begin{itemize}
    \item Many data management systems process external data that resides outside their storage for which they have little or no information. (Examples include: Apache AsterixDB \cite{asterixdb}, Apache Spark \cite{spark}, and Oracle \cite{oracle}.)
    \item The accurate sizes of join inputs may not be known if they result from other operators instead of being base relations.
    \item Newly developed DBMSs may not have statistics available until they become more mature in other dimensions.
\end{itemize}

Not having sufficient statistics can be detrimental to the performance of operators whose designs depend on such information. \cite{DBLP:conf/vldb/NakayamaK88} has proposed Dynamic HHJ to address the unbalanced join attribute values distribution by dynamically destaging the partitions at the runtime of a join query.

Investigating the Dynamic HHJ algorithm reveals several design questions that must be explored carefully, as they may impact the system's overall performance:
\begin{itemize}
    \item Number of partitions: How many partitions should the records be hashed into if the sizes of inputs are unknown or inaccurate?
    \item Partition Insertion: How can we find a "good" page (memory frame) within a partition for inserting a new record?
    \item Victim Selection Policy: How can we select a "good" partition to spill in the case of insufficient memory?
    \item Growth Policy: How many memory frames should a spilled partition be allowed to occupy?
\end{itemize}
With this motivation, this paper is an experimental survey of the trade-offs in designing a robust Dynamic HHJ algorithm. We answer the questions above through a comprehensive evaluation of different design aspects of the Dynamic HHJ algorithm and evaluate the alternative options through extensive experimental and model-based analyses.

The first contribution of this paper is to propose a lower bound and a default value for the number of partitions for Dynamic HHJ. We show that our proposed lower bound, while simple, can reduce the total amount of I/O by a factor of three in some investigated scenarios. Second, we study different partition insertion algorithms to efficiently find a frame with enough space in the target partition. We evaluate the effectiveness of these algorithms on partition compactness (fullness) and total I/O reduction. Additionally, we propose and evaluate two policies for allocating memory frames to spilled partitions. Finally, we propose and implement various dynamic destaging (victim selection) strategies and evaluate them under different scenarios such as different record size distributions, join attribute value distributions, and combinations thereof. The suggested optimization techniques and algorithm variants have been implemented in the Apache AsterixDB system and evaluated on different storage types, including HDD, SSD, and Amazon EBS.

The remainder of the paper is organized as follows: Section \ref{Background} provides background information on Apache AsterixDB and the workflow of the HHJ and Dynamic HHJ operators. Section \ref{RelatedWork} discusses previous work related to this study. In Section \ref{NumPartitions}, we discuss the lower bound on and 
the suggested default number of partitions to use in practice. Section \ref{PartitionInsertion} introduces and evaluates different partition insertion algorithms. In Section \ref{GSNGNS}, two policies for the growth of spilled partitions are discussed and evaluated. Section \ref{VSPolicies} discusses and evaluates various destaging partition selection policies. In Section \ref{Optimization}, some optimization techniques in AsterixDB are discussed before Section \ref{conclusion} concludes the paper.
\section{Background}\label{Background}
\subsection{Hybrid Hash Join}

\begin{figure*}
  \centering
  \includegraphics[scale=0.5]{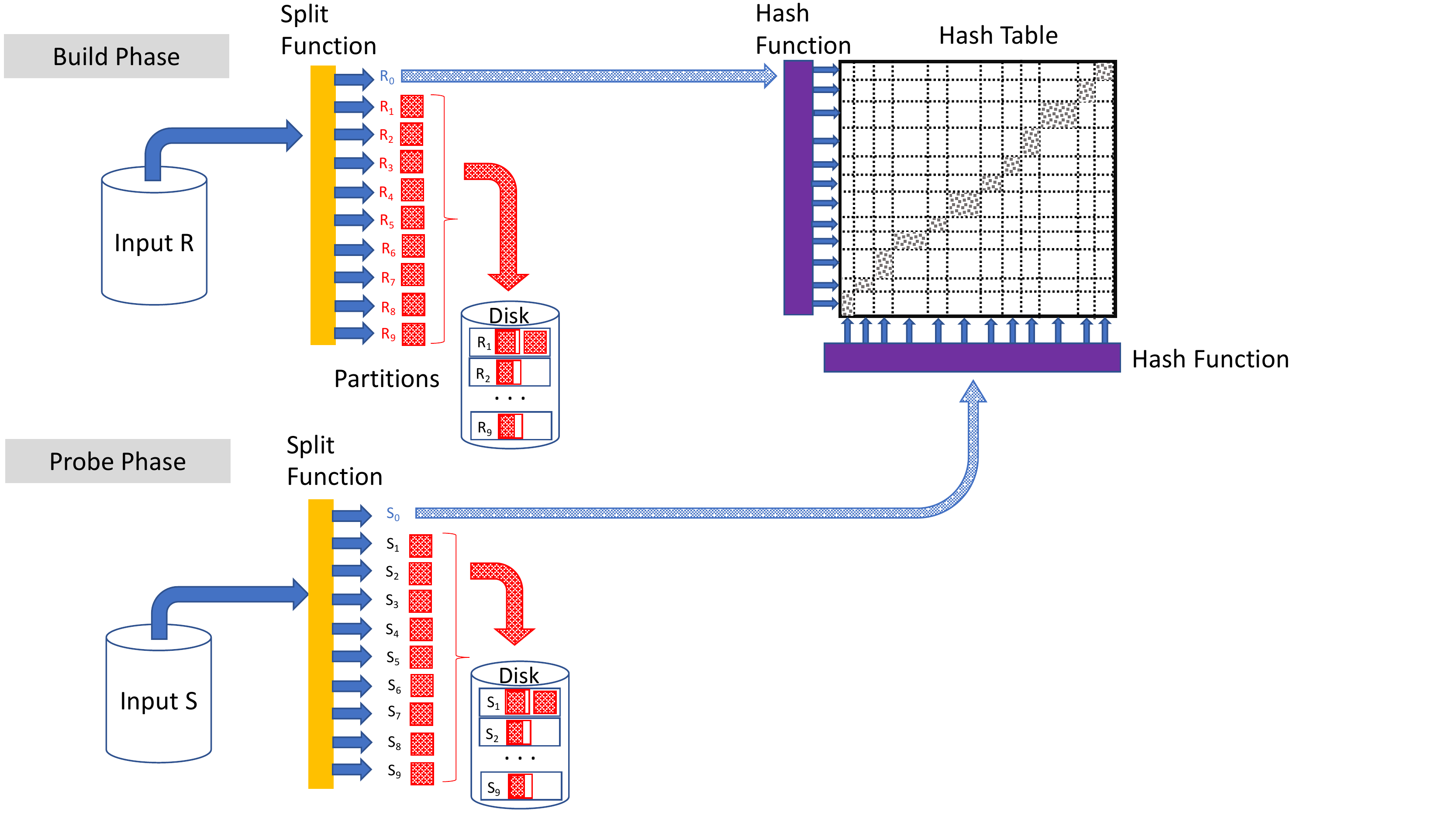}
  \caption{Workflow of Original Hybrid Hash Join}
  \label{fig:OriginalHHJ}
\end{figure*}

 Like other hash-based join algorithms, HHJ uses hashing to stage large inputs to reduce record comparisons during the join. HHJ has been shown to outperform other join types in computing equijoins of two datasets. It was designed as a hybrid version of the Grace Hash Join and Simple Hash Join algorithms \cite{DBLP:journals/tods/Shapiro86,10.1145/602259.602261}. All three mentioned hash join algorithms consist of two phases, namely "build" and "probe". During the build phase, they partition the smaller input, which we refer to as "build input", into disjoint subsets. Similarly, the probe phase divides the larger input, which we refer to as "probe input", into the same number of partitions as the build input. While all three algorithms share a similar high-level design, they differ in their details, making each of them suitable for a specific scenario.
 
Grace Hash Join partitions the build and probe inputs consecutively, writing each partition back to disk in a separate file. This partitioning process continues for each partition until they fit into memory. A hash table is created to process the join once a partition is small enough to fit in memory. Grace Hash Join performs best when the smaller dataset is significantly larger than the main memory.
 
In Simple Hash Join, records are hashed into two partitions: a memory-resident and a disk (spilled) partition. A portion of memory is used for a hash table to hold the memory-resident partition's records. Simple Hash Join performs well when memory is large enough to hold most of the smaller dataset. In Grace Hash Join, the idea is to use memory to divide a large amount of data into smaller partitions that fit into memory, while Simple Hash Join focuses on the idea of keeping some portion of data in memory to reduce the total amount of I/O, considering that a large amount of memory is available. Next, we discuss the details of the HHJ operator and compare its design with its parent algorithms.
 
Like Grace Hash Join, HHJ uses hash partitioning to group each input's records into "join-able" partitions to avoid unnecessary record comparisons. Like Simple Hash Join, HHJ uses a portion of memory to keep one of the partitions and its hash table in memory, while the rest write to disk.
Keeping data in memory reduces the total amount of I/O, and utilizing a hash table lowers the number of record comparisons. The overall of Hybrid Hash Join is shown in Figure \ref{fig:OriginalHHJ}.
 
As mentioned earlier, the HHJ operator consists of two consecutive phases of build and probe. During the build phase, the records of the smaller input are scanned and hash-partitioned based on the values of the join attributes. We call the hash function used for partitioning a "split function." The records mapped to the memory-resident partition remain in memory, while the rest of the partitions are written (frame by frame) to disk. Pointers to the records of the memory-resident partition are inserted into a hash table at the end of the build phase.
 
After the build phase ends, the probe phase starts by scanning and hash-partitioning the records of the larger input. The same split function used during the build phase is used for this step. The records that map to the memory-resident partition are hashed using the same hash function used in the build phase to probe the hash table. All other records belong to spilled partitions and are written (frame by frame) to that partition's probe file on disk.

After all records of the probe input have been processed, the pairs of spilled partitions from the build phase and probe phase are processed as inputs to the next rounds of HHJ.

\subsection{Apache AsterixDB}
Apache AsterixDB \cite{asterixdb,DBLP:journals/pvldb/AlsubaieeAABBBCCCFGGHKLLOOPTVWW14,DBLP:journals/spe/KimBBBBCHJJLLMP20} is an open-source, parallel, shared-nothing big data management system (BDMS) built to support the storage, indexing, modifying, analyzing, and querying of large volumes of semi-structured data. 

The unit of data that is transferred within AsterixDB, as well as between AsterixDB and disk is called a "frame". A frame is a fixed-size and configurable set of contiguous bytes. AsterixDB uses Dynamic HHJ, whose design and optimization is the main topic of this paper. AsterixDB supports different join algorithms such as Block Nested Loop Join, Dynamic HHJ, Broadcast Join, and Indexed Nested Loop Join. However, Dynamic HHJ is the default and primary join type in AsterixDB for processing equi-joins due to its superior performance.

AsterixDB currently does not support statistics, so users may provide hints to guide AsterixDB at execution time by selecting an alternative type of join operator or by providing dataset size information. For example, a user may use the Indexed Nested Loop Join hint to request this join algorithm instead of a Dynamic HHJ. AsterixDB follows this hint whenever possible; otherwise, it utilizes Dynamic HHJ (by default). In addition, a hint to use a Broadcast Join might be advantageous when the build dataset is small enough to be sent to all nodes instead of using hash partitioning.
\begin{figure*}
  \centering
  \includegraphics[scale=0.5]{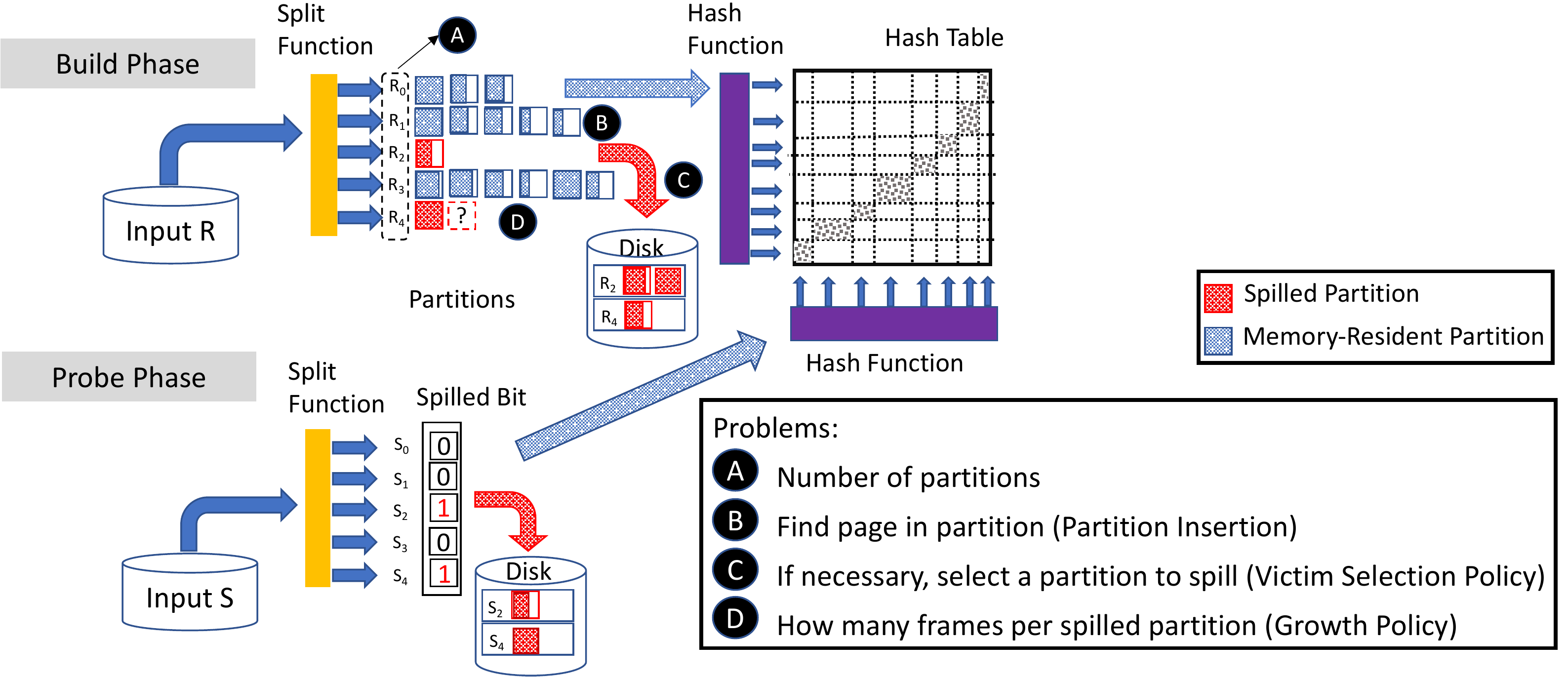}
  \caption{Workflow of Dynamic HHJ}
  \label{fig:DynamicHybridHashJoin}
  \vspace{-4mm}
\end{figure*}
The current release of AsterixDB follows the join order in a query’s FROM clause for determining the build and probe inputs. The first input in the FROM clause will serve as the probe relation; the rest will be build inputs. 

We chose Apache AsterixDB as our primary platform for implementing and evaluating our proposed techniques for several reasons. First, it is an open-source platform that allows us to share our techniques and their evaluations with the community. More importantly, AsterixDB is a parallel big data management system for managing and processing large amounts of semi-structured data with a declarative language. Finally, its similarity in structure and design to other NoSQL and NewSQL database systems and query engines makes our results and techniques applicable to other systems as well.

\subsection{Dynamic Hybrid Hash Join}
Dynamic HHJ was first introduced in \cite{DBLP:conf/vldb/NakayamaK88}, where the authors used dynamic destaging instead of the static predefined memory-resident partition method. 
As Figure \ref{fig:DynamicHybridHashJoin} shows, all partitions at the build phase have an equal chance to grow as long as enough memory frames are available. Each partition uses an array to hold its in-memory frames. This flexibility in acquiring frames may cause some partitions to receive more frames than others if join attribute values are skewed. In this case, Dynamic destaging is useful since the decision of choosing the spilling partition is made at runtime. It is also useful when the build input size or the distributions of join attribute values are unknown or inaccurate.


If memory is insufficient, one or more partitions may spill to disk to free some memory space for the incoming records. After partitioning the build dataset's records, the pointers to the records of the surviving memory-resident partitions are hashed and inserted into the hash table to be probed. The spilled partitions are processed one by one in the next rounds of the Dynamic HHJ operator.

\section{Related Work}\label{RelatedWork}
HHJ was first proposed in \cite{10.1145/602259.602261}. It was shown to have superior performance compared to other types of join using simple cost models, especially if a large amount of memory is available\cite{DBLP:journals/tods/Shapiro86}. 
In \cite{DBLP:journals/vldb/HaasCLS97}, the authors provided a more detailed cost model to determine the optimal buffer allocation for various join types.

One of the key problems in configuring HHJ for execution is to choose the number of partitions into which to hash the records.
In \cite{DBLP:journals/tods/Shapiro86}, the author provided an equation for calculating the number of partitions based on the memory and build input size.
In \cite{DBLP:conf/vldb/KitsuregawaNT89}, the authors derived an upper bound on the number of partitions and then merged smaller partitions to reduce the fragmentation in each partition, which is helpful when the join attribute values are skewed. Our paper introduces a lower bound and a default value for the number of partitions and shows how it can significantly reduce the total amount of I/O.

Another challenge for executing HHJ is to efficiently find a frame with sufficient space in the target partition for each incoming record. This problem is similar to the Bin-Packing problem \cite{DBLP:journals/ipl/Liang80,DBLP:conf/icalp/DosaS14}. 
The problem has also been widely studied in the operating system and the DBMS literature \cite{DBLP:conf/sigmod/McAuliffeCS96,DBLP:journals/jsa/WezenbeekW93} for managing free disk space. This paper will examine those algorithms and a few more for inserting records in partitions during HHJ. The difference between our work and disk-related studies is that in our work records will not reside in the partitions long term, and no deletion apart from partition spilling happens in this case.

The authors of \cite{DBLP:conf/vldb/NakayamaK88} proposed a dynamic destaging scheme where the partition written to disk is selected dynamically during execution. In \cite{DBLP:conf/vldb/GraefeBC98}, Graefe et al. detailed the optimization techniques and the design of Dynamic HHJ variant in Microsoft SQL Server. Those two studies are closely related to our work; both choose the largest partition to be written to disk. Despite some reasoning, the authors discuss no other options, nor do they evaluate them. Our study defines 13 different possibilities and evaluates them under various record sizes and join attribute value distributions.

Regarding AsterixDB \cite{asterixdb,DBLP:journals/pvldb/AlsubaieeAABBBCCCFGGHKLLOOPTVWW14,DBLP:journals/spe/KimBBBBCHJJLLMP20}, the details of its default Dynamic HHJ can be found in \cite{DBLP:journals/spe/KimBBBBCHJJLLMP20}.


\section{Number Of Partitions}\label{NumPartitions}
The first step in configuring the HHJ operator is to determine the number of the partitions for partitioning the input datasets. The purpose of this section is twofold: 
\begin{enumerate*}
\item Choosing the number of partitions for the cases where no a priori information about input datasets is provided.
\item Providing a lower bound on the number of partitions to prevent excessive spilling due to inaccuracy of the provided information. 
\end{enumerate*}

There are two main constraints to be considered when choosing the number of partitions:
\begin{enumerate*}
\item An HHJ operator needs at least two partitions to divide the input dataset into smaller subsets.
\item Each partition needs at least one output frame in order not to spill less than half-full frames to disk.
\end{enumerate*} 

As such, the number of partitions for an HHJ should be chosen from the range of:
\begin{equation}\label{eq:partitionBound}
Number \,of \,Partitions = \left[2,\, \#\, of\, memory\, frames\right]
\end{equation}
In \cite{DBLP:journals/tods/Shapiro86}, the author offers the following equation to calculate the number of partitions for an HHJ operator.

\begin{equation}\label{eq:shapiro}
B = \ceil{\frac{|R|*F-|M|}{|M|-1}}
\end{equation}

|R| represents the size of the build input in frames, F is a fudge factor, |M| represents the size of the memory in frames available to this join operator, and B is the number of disk-resident partitions. Based on this equation, the HHJ operator will use B+1 partitions (including a memory-resident partition) and finish in B+1 rounds. 

While this equation calculates the number of partitions in a way that minimizes the total rounds in HHJ and thus reduces unnecessary I/O, any inaccuracy in estimating its input parameter, |R|, might drastically impact the operator’s performance. This is especially true when only a few partitions are created (large memory). In this case, data is distributed among just a few partitions, causing a high penalty for spilling a partition as a large amount of data will be written to disk.

\begin{figure}[!h]
  \centering
  \begin{minipage}[b]{0.47\textwidth}
    \includegraphics[width=\textwidth]{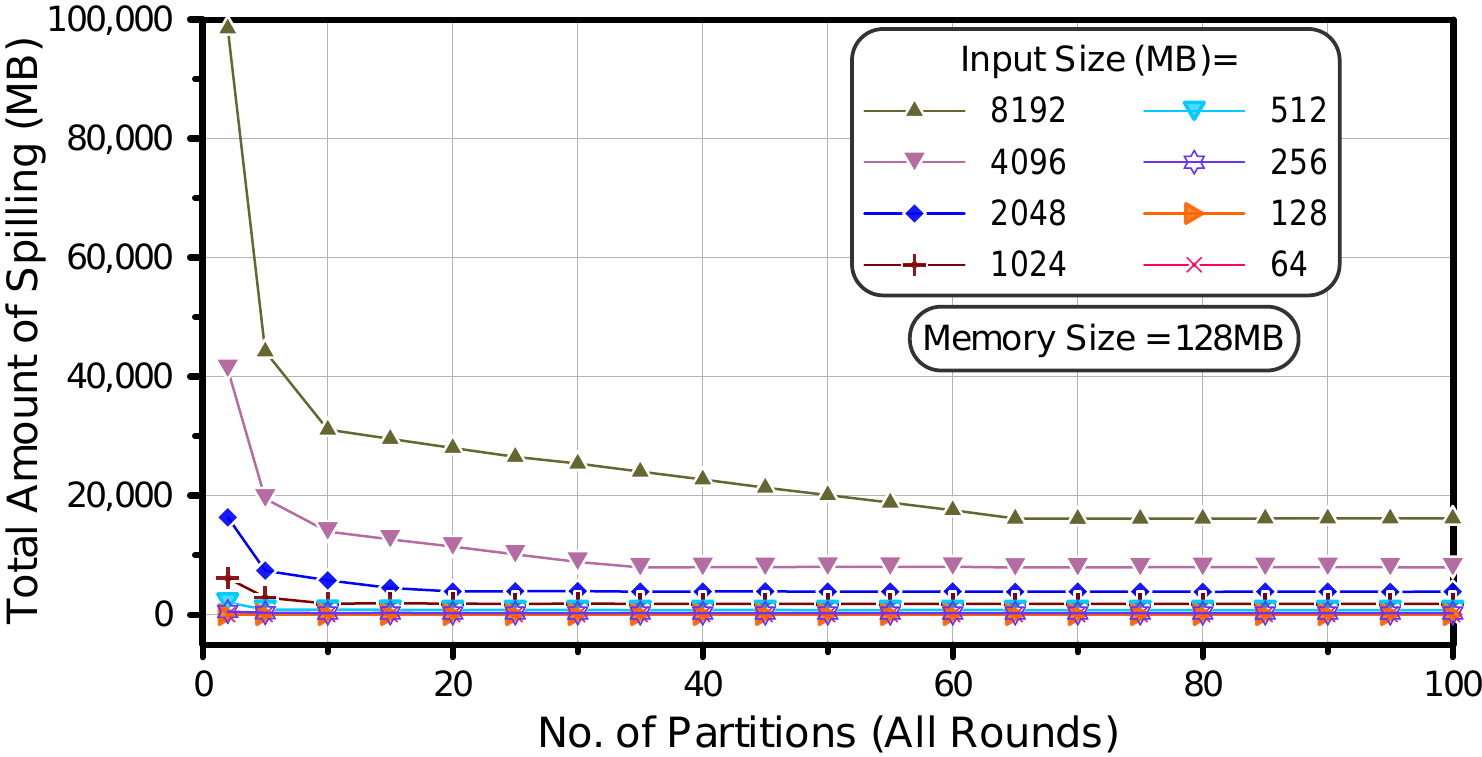}
     \caption{Impact of Number of Partitions (Fixed for All Rounds) on the Total Amount of Spilling}
      \label{fig:numOfPartitions_fixed}
      \vspace{-1mm}
  \end{minipage}
  \hfill
  \end{figure}

Figure \ref{fig:numOfPartitions_fixed} shows the result of a simulation study that explores the impact of the number of partitions on the amount of data written to disk during the execution of an HHJ operator.  Final result writing is excluded from this measurement. We use the same number of partitions for all rounds of HHJ in this experiment. Both the build and probe inputs contain the same size of data for simplicity. 
  The amount of available memory stays fixed at 128 MB during this simulation. At the same time, the sizes of the inputs change to cover all the cases from when the build dataset fits in memory to when it is 64 times larger than memory. As Figure \ref{fig:numOfPartitions_fixed} shows, the number of partitions does not impact the amount of spilling significantly if a large portion of data fits in memory (input sizes less than or equal to 2048 MB in this example); however, this is not true when the data size is considerably larger than the memory size (input sizes equal to or larger than 4096 MB in this example). In the latter case when data size significantly exceeds the memory size (input sizes equal to or larger than 4096 MB in Figure \ref{fig:numOfPartitions_fixed}), choosing a small number of partitions leads to a handful of large-sized partitions causing extra rounds of HHJ and large amount of spilling to disk. On the other hand, while using a larger number of partitions can reduce the total amount of spilling, it can make the join’s I/O pattern more random due to frequent writings of partitions containing just a few frames.
  \begin{figure}[!h]
{\small    \begin{minipage}[b]{0.45\textwidth}
    \centering
    \begin{tabular}{|l | l || l|  l |c|c|c|c| } 
\hline
\thead{Build Size (MB)} & \thead{\# Partitions} &\thead{Build Size (MB)} & \thead{\# Partitions} \\
\hline
64 & 2 & 1024 & 10 \\
128 & 2 & 2048 & 20 \\
256 & 2 & 4096 & 41 \\
512 & 5 & 8192 & 83 \\
\hline
\end{tabular}
\captionof{table}{Number of partitions calculated by Equation \ref{eq:shapiro}}
\label{table:numOfPartitions}
\vspace{-5mm}
\end{minipage}}
\end{figure}

 \begin{figure}[!h]
  \centering
  \begin{minipage}[b]{0.47\textwidth}
    \includegraphics[width=\textwidth]{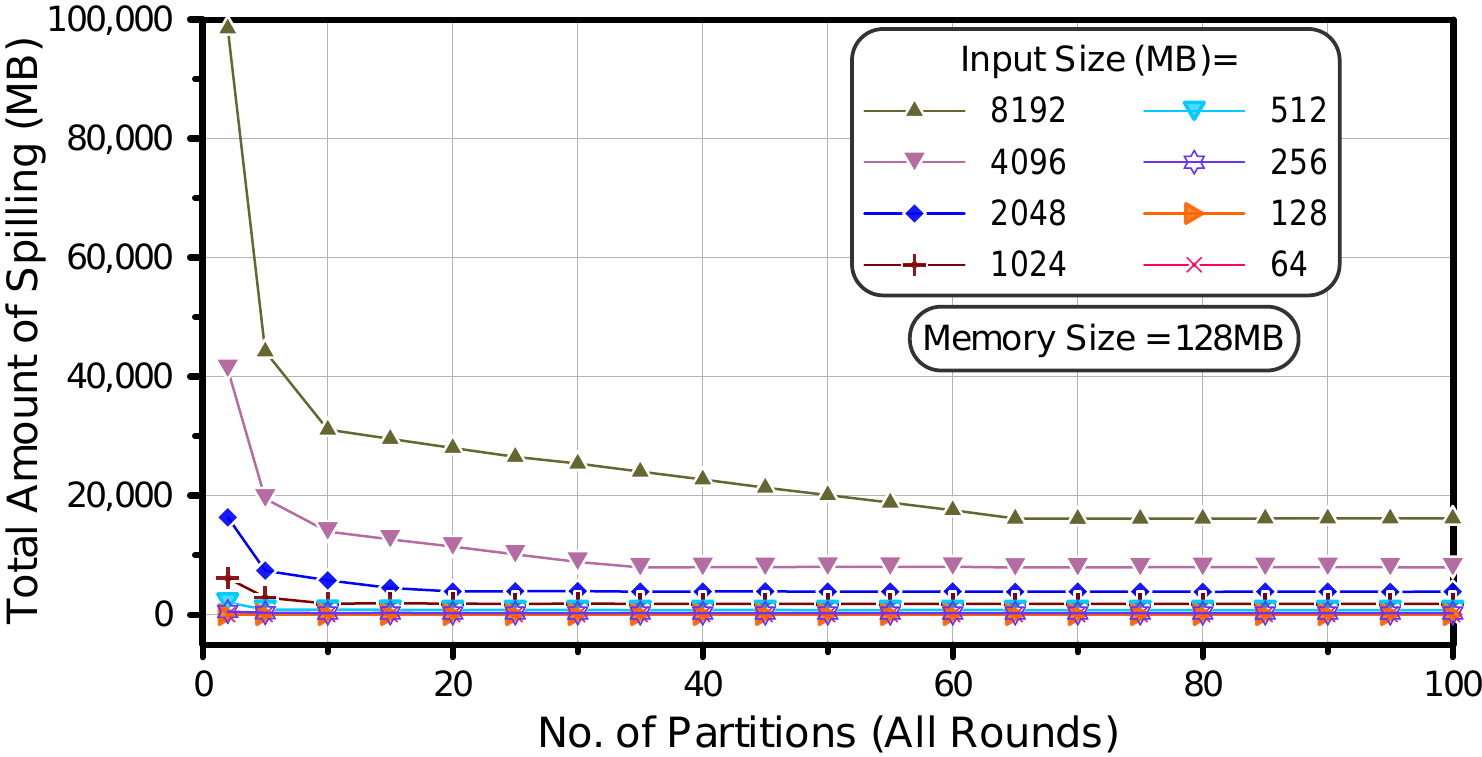}
     \caption{Impact of Number of Partitions (First Round Only) on the Total Amount of Spilling}
      \label{fig:numOfPartitions_variable}
      \vspace{-3mm}
  \end{minipage}
  \hfill
  \end{figure}
Fragmentation within frames is another downside of having a large number of partitions. In \cite{DBLP:conf/vldb/KitsuregawaNT89}, the authors defined an upper bound for the number of partitions in order to reduce fragmentation and random writes due to too many single-frame partitions. However, to the best of our knowledge, no lower bound on the number of partitions has been suggested to improve the performance of the HHJ algorithm. 
Table \ref{table:numOfPartitions} shows the number of partitions calculated using Equation \ref{eq:shapiro} given accurate inputs.

We can use Equation \ref{eq:shapiro} to calculate the number of partitions for the next rounds of HHJ as the sizes of spilled partitions are known. Figure \ref{fig:numOfPartitions_variable} shows that how using the spilled partition sizes in accurately calculating the number of partitions for the next rounds of an HHJ can reduce the total amount of spilling of this operator. 
  \begin{figure}[!h]
	\centering
	\begin{minipage}[b]{0.47\textwidth}
		\includegraphics[width=\textwidth]{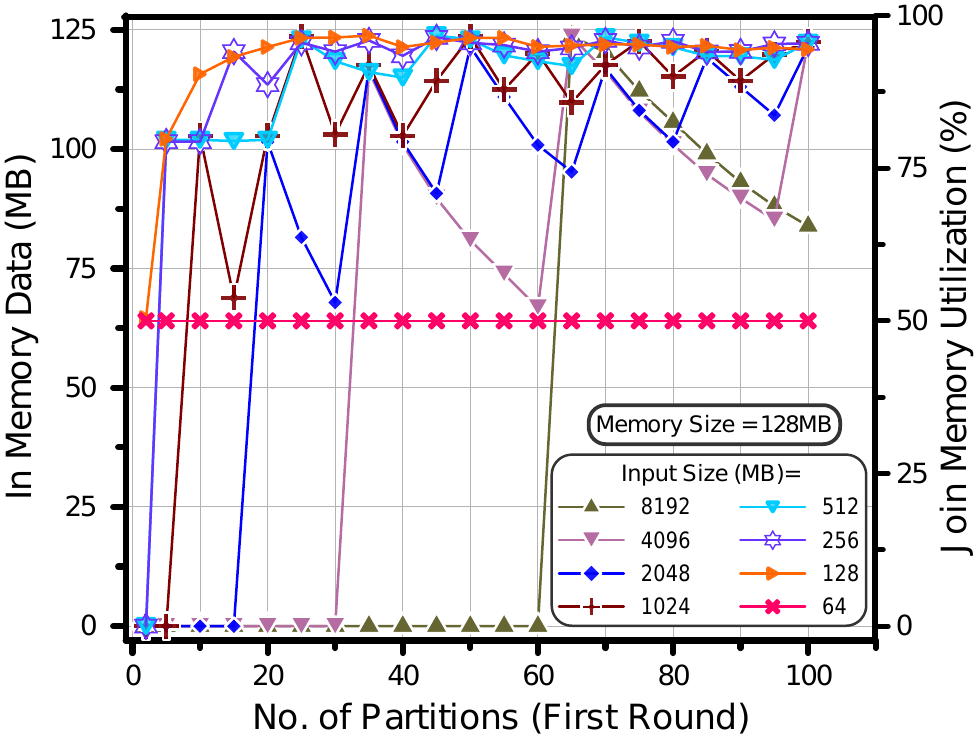}
		\caption{Impact of Number of Partitions on the Amount of Data Remaining in Memory}
		\label{fig:numOfPartitions_inMemoryData}
		\vspace{-3mm}
	\end{minipage}
\end{figure}

\noindent Figure \ref{fig:numOfPartitions_inMemoryData} shows the impact of the number of partitions on the amount of build input data that remains in memory in the first round of HHJ. As this figure shows, if we drew a vertical line at $20$ partitions on the x-axis, we would see that most of the lines have utilized more than $78\%$ of their given memory. As a result, increasing the number of partitions any further can only slightly increase their amount of in-memory data. The only exceptions are the 4096 MB and 8192 MB input sets. For these cases $20$ is still a good choice for the \textit{minimum} number of partitions (lower bound) since as Figure \ref{fig:numOfPartitions_fixed} shows, the amount of spilling lowers by having $20$ or more partitions.

The reason that the lines in Figure \ref{fig:numOfPartitions_inMemoryData} follow an up-and-down pattern is that changing the number of partitions impacts the partitions’ sizes. In cases where the lines follow an uphill trend (more in-memory data), the same number of partitions spills as past; however, less overspilling occurs because the size of each partition is reduced by increasing the number of partitions. After some point, the amount of in-memory data in each partition is decreased by increasing the number of partitions; causing more partitions to spill in order to release enough memory for incoming records. Spilling more partitions causes overspilling, the impact of which is seen in Figure \ref{fig:numOfPartitions_inMemoryData} when the lines follow a downhill trend.

To summarize, we recommend $20$ as a good default choice for the number of partitions when accurate information on the input sizes is unavailable (first round of HHJ in some cases) and as the minimum number of partitions in all other cases. As Figures \ref{fig:numOfPartitions_fixed} and \ref{fig:numOfPartitions_variable} show, the amount of I/O drops significantly before $20$ partitions, as most of the lines are flat before this point. This makes $20$ a reasonable choice for the number of partitions. More generally, by having a minimum of $20$ partitions, each spilled partition spills no more than {5\%} of the data, so the potential for significant “spilling error” is low. Lastly, $20$ partitions does not cause too many random I/Os since data will be written to only a few (at most $20$) files on the disk. A modest filesystem cache can turn many of these random writes to sequential ones (Elevator Algorithm).

\section{Partition Insertion}\label{PartitionInsertion}
After choosing the number of partitions (P) (Figure \ref{fig:DynamicHybridHashJoin} - A), the build phase starts by reading its input into memory one frame after another. The split function is applied to each incoming record's join attribute(s) to find their destination partition. Next, we need to search for a frame with enough space to hold the record in the destination partition's array of in-memory frames (Figure \ref{fig:DynamicHybridHashJoin} - B). The search starts from the newest allocated frame and proceeds towards the oldest one. If this search is unsuccessful, a new frame will be allocated and appended to this partition's in-memory frames array if enough memory is available. However, if the available memory is not sufficient for a new frame allocation, one of the memory-resident partitions will be selected for spilling to release some memory space. This choice is called victim selection and will be discussed in Section \ref{VSPolicies}. \newline
\textbf{Problem Definition.} Our goal for partition insertion is to make each partition as non-fragmented as possible by choosing the destination frame for each incoming record in such a way that minimizes the free space in each frame. Otherwise, under-filled frames can lead to extra I/O and additional rounds of HHJ, which can negatively impact the execution time of queries and the system's throughput. On the other hand, searching for a proper frame for each record could be CPU-time-consuming, depending on the search strategy and the number of frames searched. Our goal is to find a destination frame efficiently while making the partition as compact as possible. Two influential factors should be considered for designing partition insertion algorithms. First, there will be no record deletions to cause fragmentation in this scenario; only a complete partition will be written to disk in case of insufficient memory. 
Second, records can come in many different sizes. This variation in record sizes adds to the complexity of partition insertion for two reasons. First, the space required for each record is different from other records. Second, the insertion of variable-sized records in fixed-size frames leaves a different amount of free space in each frame. Placing variable-sized objects in a fixed-size space is known as an "online object placement" or "online organization" problem. It is an example of the online bin-packing problem, a well-known NP-hard problem. Some object placement strategies have been studied and optimized for free space management on disk for permanent placement of objects \cite{DBLP:conf/sigmod/McAuliffeCS96}; however, they may not exhibit similar performance characteristics when used for memory space management. 

In the following, we present different algorithms for partition insertion and evaluate them under different data distributions and for different storage devices. The algorithms considered here are: 

\textbf{Append($n$).} Append($n$) performs a search on the last $n$ frames of the target partition in the order of the newest frame to the oldest. The incoming record will be placed in the first frame with enough space. If no such frame is found, a new frame with enough space will be appended to this partition.\newline
\textbf{First-Fit.} In the First-Fit algorithm, the search starts from the last (newest) frame towards the first (oldest) frame of the partition and stops as soon as a frame with enough space for the record is found. In First-Fit's worst-case scenario, all frames are searched and a new frame is appended upon an unsuccessful search. \newline%
\textbf{First-Fit($\%p$)}. This is a parameterized and more general version of the First-Fit algorithm in which at most $\%p$ of the partition's frames are searched for the record insertion. Similar to the previous algorithms, the search proceeds from the newest frame towards the oldest. It appends a new frame to the array and inserts the record if no frame with enough space is found. In comparison to First-Fit, this algorithm provides a better balance between extensive search and the compactness of the frames in the partition. This algorithm is similar to Append($n$) as they both start searching from the end of the frames array and stop if the stopping criteria are met. The stopping criteria in Append($n$) is $n$ frames, while in First-Fit($\%p$) it is $\%p$ of frames.\newline
\textbf{Best-Fit.} Best-Fit, a well-known space management algorithm, searches through all of the partition's frames to find the frame with the smallest free space that can accommodate the record. This algorithm tries to maximize frame compactness based on the current state of the frames and the size of the record being inserted.\newline
\textbf{Next-Fit.} Next-Fit starts searching from a different location for each record to avoid checking some frames over and over again. In this algorithm, the search is guided based on the size and insertion location of the previous record.\newline
As a modified version of the First-Fit algorithm, Next-Fit initially starts searching from the end of the partition's array. However, after the first record, the search starts from the location where the previous record was inserted. If the size of the current record is larger than the previous record, the search continues toward the newer frames. However, if the current frame is smaller than the previous record, the older frames are searched first. In the latter case, if no frame with enough space is found, then the search continues toward the newer frames. A new frame is appended to the end of the partition's frame array if no frame with enough space is found.\newline
\textbf{Random($\%p$)} In this algorithm, for each record, up to $\%p$ of the partition's frames are randomly searched. The search stops as soon as a frame with enough space is found. This algorithm avoids searching the same frames extensively and unnecessarily by its random selection of frames. We tried different random number generators such as Java's default random number generator, Mersenne Twister Fast \cite{Mersenne}, C++ 11 MinSTD, and XorShift 64 bits and compared their performance to choose the least expensive random number generator for our case. Our experiments showed that these random number generators performed very similar to each other. As such, we decided to use Java's default random number generator.

\subsection{Choosing the best parameter values}
As we discussed, some of the partition insertion algorithms such as Random($\%p$), Append($n$), and First-Fit($\%p$) have a parameter that needs to be properly set. We compared the performance of these algorithms under different value settings for their parameters using the 1 Large Record Coexist setting whose specification can be found in Table \ref{table:datasetspec}.

\begin{figure*}
  \centering
  \includegraphics[scale=0.8]{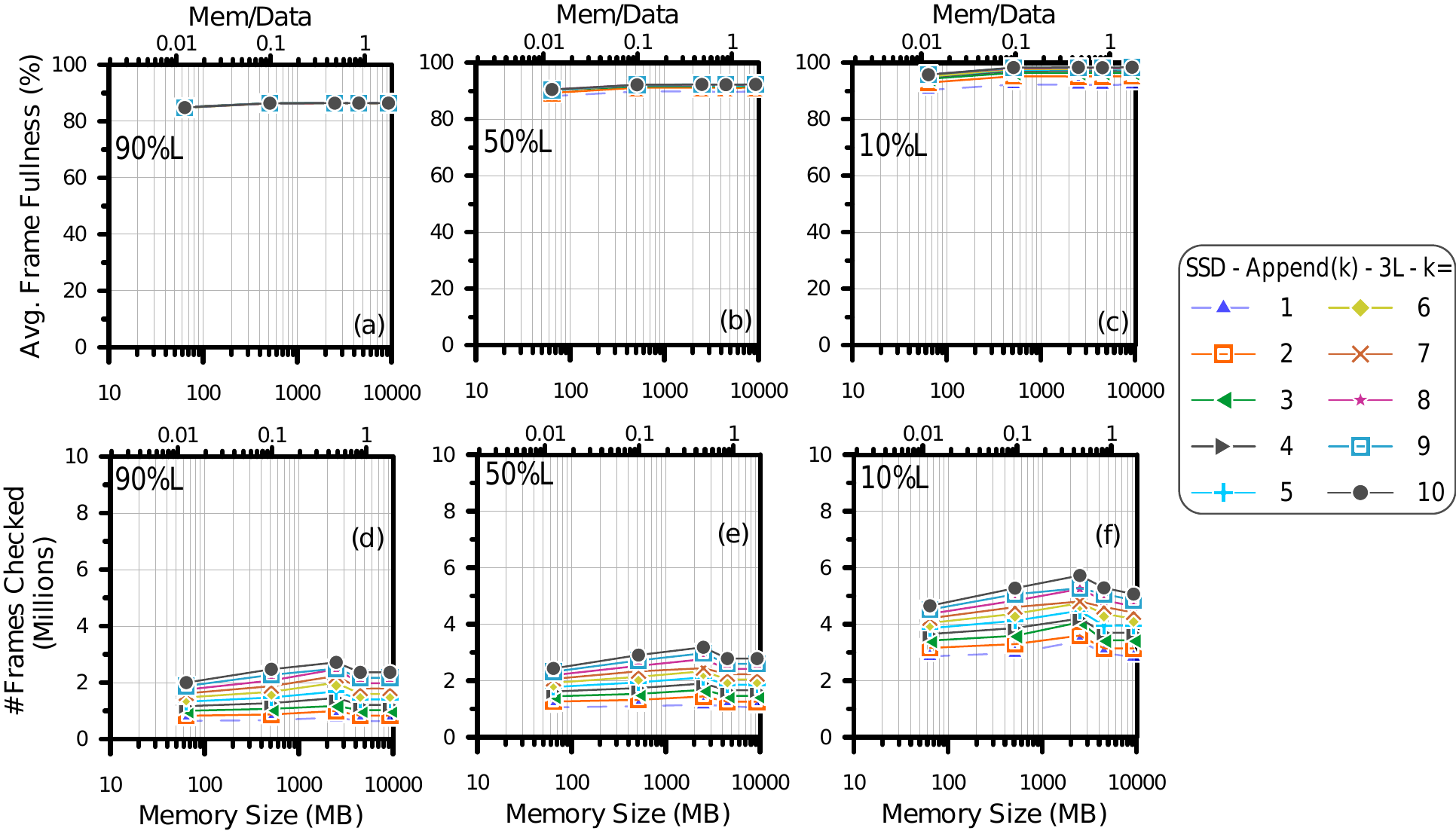}
  \caption{Choosing The Best Parameter Value for Append(K) (1 Large Record Coexist). (a) - Average frame fullness when $90\%$ of records are large. (b) - Average frame fullness when $50\%$ of records are large. (c) -Average frame fullness when $10\%$ of records are large. (d) - Total number of searched frames when $90\%$ of records are large. (e) - Total number of searched frames when $50\%$ of records are large. (f) - Total number of searched frames when $10\%$ of records are large.}
  \label{fig:appendk}
\end{figure*}

\begin{figure*}
  \centering
  \includegraphics[scale=0.8]{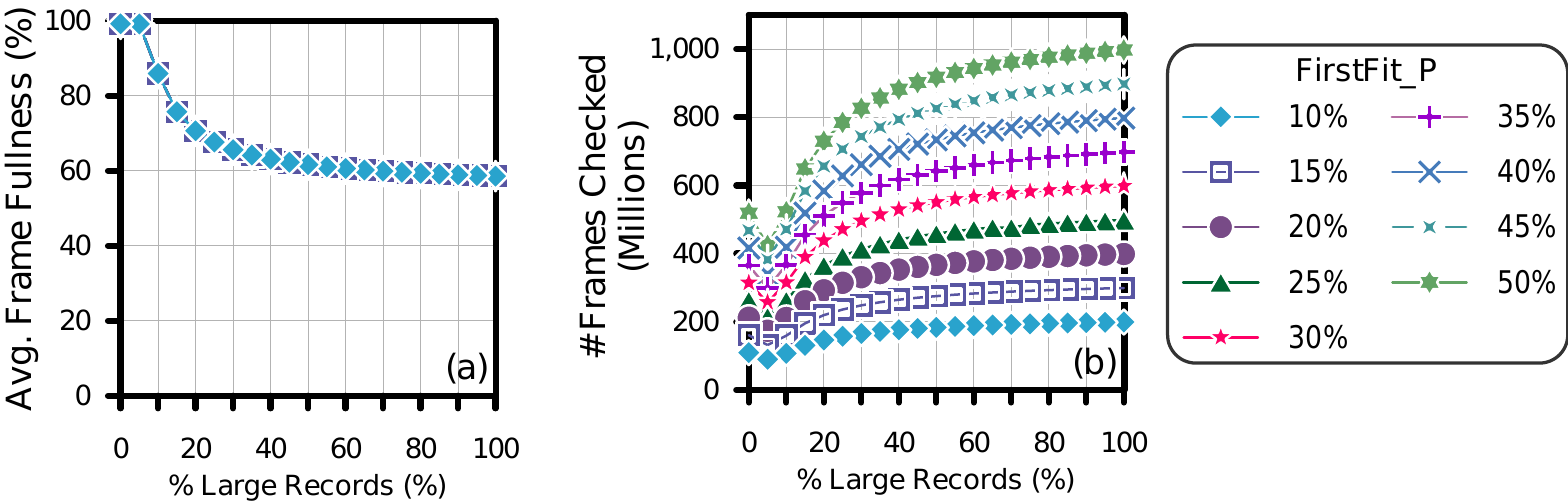}
  \caption{Choosing The Best Parameter Value for First-Fit(P) (1 Large Record Coexist). (a) - Average frame fullness by different parameters of First-Fit(P). (b) - Number of searched frames by different parameters of First-Fit(P).}
  \label{fig:firstfitp}
\end{figure*}

\begin{figure*}
  \centering
  \includegraphics[scale=0.8]{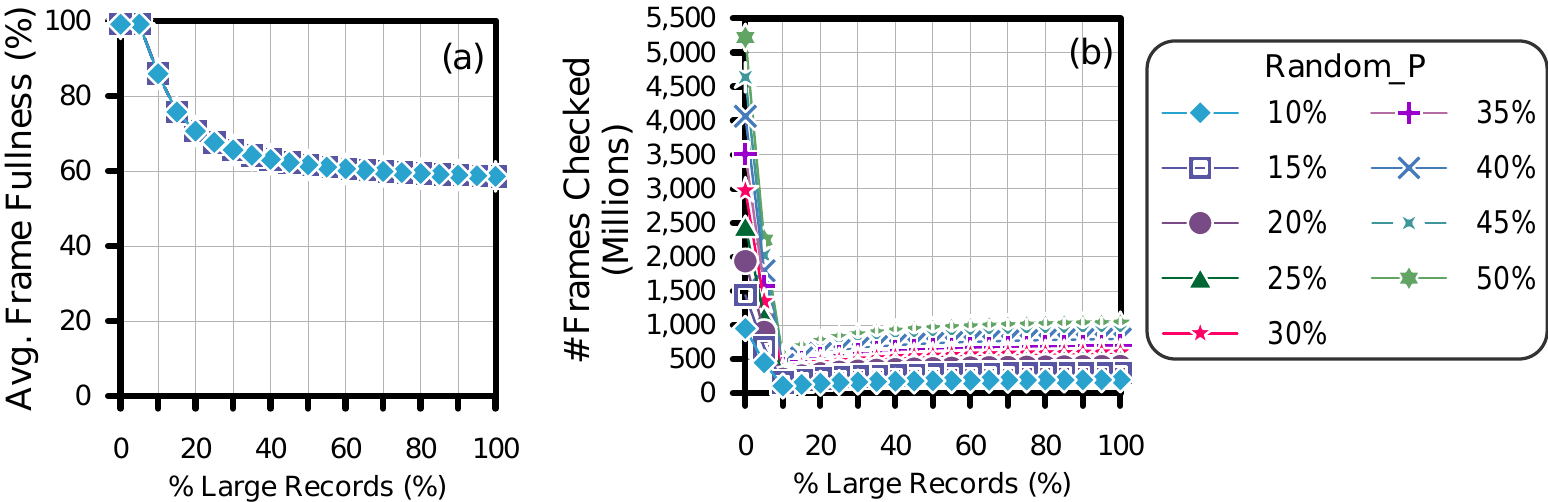}
  \caption{Choosing The Best Parameter Value for Random(P) (1 Large Record Coexist). (a) - Average frame fullness by different parameters of Random(P). (b) - Number of searched frames by different parameters of Random(P).}
  \label{fig:randomp}
\end{figure*}

Figures \ref{fig:appendk}-a, \ref{fig:appendk}-b, and \ref{fig:appendk}-c show, on average, how much of the frames are filled with records when $90\%$, $50\%$, and $10\%$ of the records are large. As we can see, all of the different parameters lead to a similar frame fullness in the $90\%$ and $50\%$ cases as the majority of the records are large, only one large record can fit in a frame, and there are few small records to fill the holes in frames. However, when $10\%$ of the records are large, these parameters' fullness results slightly differ from one another. Append($8$) appears to have a frame fullness close to the frame fullness of Append($9$) and Append($10$); however, as Figures \ref{fig:appendk}-d, \ref{fig:appendk}-e, and \ref{fig:appendk}-f show, Append($8$) checks fewer frames than Append($9$) and Append($10$). 
Figure \ref{fig:firstfitp} and Figure \ref{fig:randomp} show the average frame fullness and the number of searched frames for different parameter values for First-Fit(P) and Random(P). In this experiment, enough memory is available to keep all of the joins in memory. As Figure \ref{fig:firstfitp}-a and Figure \ref{fig:randomp}-a show, all parameters of First-Fit(P) and Random(P) have a similar average frame fullness; however, they differ in the number of frames that they search. Hence, based on these experiments, Random($\%10$), Append($8$), and First-Fit($\%10$) were found to achieve the highest degree of frame fullness with the least number of frames being checked. We will therefore study just these settings as we move forward with these policies.

\subsection{Dataset and Experiment Design}

 We use an updated and modified version of the Wisconsin Benchmark \cite{DBLP:books/mk/gray91/DeWitt91} data to evaluate the partition insertion algorithms. Its attributes and datasets' high tunability and selectivity make the Wisconsin Benchmark's dataset a good synthetic benchmark dataset for evaluating and benchmarking join queries. The range of integer-based attributes and their relation with other integer attributes, the efficiency in generating random and unique integers, and the provision of strings with extendable lengths are some of the features that make it possible to tune the datasets as needed.
 
 We use variable-length records, one of the modifications added to the Wisconsin Benchmark data in \cite{DBLP:conf/sigmod/Jahangiri21}, to introduce two groups of small-sized and large-sized records with a specific ratio between these two groups. We use what we call the 1-Large Record Coexist, 3-Large Record Coexist, and All Small Records datasets in this study, each of which is 1 GB in size. The names of 1-Large Record Coexist and 3-Large Record Coexist come from the number of large records that can fit in one frame. Table \ref{table:datasetspec} contains the details of the mentioned datasets.
 {\small{ \begin{table}[hbt!]
    \centering
    
 \begin{tabular}{|l|l|l| }
\hline
\thead{Dataset} & \thead{Small Records} & \thead{Large Records}  \\
\hline
1-Large Record Coexist & 700 B - 1500 B & 18 KB - 20 KB   \\
3-Large Records Coexist & 700 B - 1500 B & 8 KB - 10 KB\\
All Small Records & 700 B - 1500 B & None\\
\hline
\end{tabular}
\captionof{table}{Dataset Specifications}
\label{table:datasetspec}
\vspace{-10mm}
\end{table}}}

\begin{figure*}
  \centering
  \includegraphics[scale=0.70]{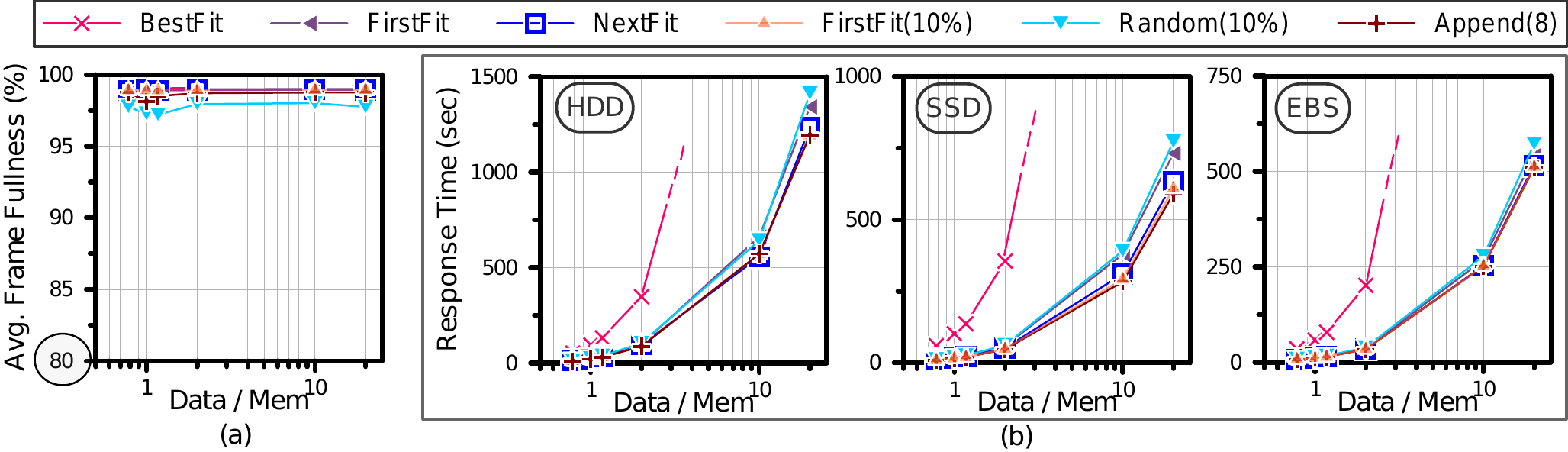}
  \caption{Partition Insertion - Small Record Sizes (a) Average frame fullness (b) Response time on different storage types}
  \label{fig:DI_smallRecords}
\end{figure*}

\subsection{Evaluation of Partition Insertion Algorithms}
This section evaluates the performance of the described partition insertion algorithms for two cases: 
\begin{enumerate*}
    \item Records are all small and similar in size.
    \item Records belong to two groups of small and large records where specific ratios between these two groups are defined.
\end{enumerate*}

\subsubsection{\textbf{Small Records Experiment}}
In our first experiment, we use a single join query to evaluate the performance and efficiency of the partition insertion algorithms when records are small and similar in size. Both the build and probe datasets are 1GB in size and follow the All Small Records dataset configuration. Each memory frame is 32KB and thus can hold between 21 to 65 records. In this experiment, we are interested in comparing the partition insertion algorithms with respect to the average frame fullness (compactness) and the query response time to reflect on the efficiency of each algorithm in reaching this degree of frame fullness. For simplicity, enough available memory is provided to avoid any possible spilling.

Figure \ref{fig:DI_smallRecords}(a) shows the average frame fullness as a function of the ratio of the build dataset size to the amount of available memory. As this figure shows, all algorithms deliver a high and similar average frame fullness when the records are small. This is because small records can easily fit in most frames and increase the average frame fullness by minimizing the leftover space in each frame. 

Next, we analyze the performance of the different partition insertion algorithms in reaching their reported frame compactness. Figure \ref{fig:DI_smallRecords}(b) exhibits the response time of the partition insertion algorithms for three storage types of HDD, SSD, and Amazon EBS.

The similarity in the size of the records makes the frames, especially the older ones, similarly full. Additionally, suppose a previous record could not find a frame by checking all of the partition's frames due to similarity in record sizes. In that case, it is likely that the next record will not fit in those frames either. Therefore, algorithms that start from the newest allocated frame and exhaustively search for a fitting frame, such as First-Fit and Best-Fit, will have a higher response time due to more and unnecessary searches. This extra work can significantly impact the system's performance if the datasets consist of a large number of small records, as this search is done per record. However, the algorithms with a varying starting point of search, such as Next-Fit and variations of Random, have a better chance to find an accommodating frame quickly. Additionally, those algorithms with predefined stopping criteria such as First-Fit($10\%$), Append($8$), and Random($10\%$) avoid unnecessary searching by giving up early and allocating a new frame.

As Figure \ref{fig:DI_smallRecords}(b) shows, the CPU cost due to extensive searching in Best-Fit significantly degrades its performance in all three storage types. Random($10\%$) is the second-worst algorithm with a slightly higher response time than the others. Although Random($10\%$) benefits from the additional stopping criteria, the high time-overhead of the Random function and the high frequency of calling it degrades its performance. First-Fit is the third-worst algorithm in our experiments. First-Fit has a higher response time than the algorithms with a guided search method (Next-Fit) or additional stopping criteria. This is due to the extensive search of First-Fit. However, the performance of First-Fit is much better than Best-Fit, another extensive search algorithm, as First-Fit stops if it finds a suitable frame. This "first find" strategy has a high impact, especially in this experiment, as all of the records are small and have a good chance to fit in even a relatively full frame. Next-Fit and First-Fit($10\%$) perform similarly here with relatively low response times. Next-Fit's different starting point and its guided search improve its performance. The early termination due to stopping criteria in First-Fit($10\%$) makes it one of the best-performing algorithms here. Append($8$), however, seems to be the best algorithm in this experiment. As Figures \ref{fig:DI_smallRecords}(a) and \ref{fig:DI_smallRecords}(b) show, Append($8$) reaches a similar average frame fullness as the other alternatives with the least amount of search effort. (For each record, at most $8$ frames are checked.)

\subsubsection{\textbf{Variable Size Records}}
\begin{figure*}
  \centering
  \includegraphics[scale=0.70]{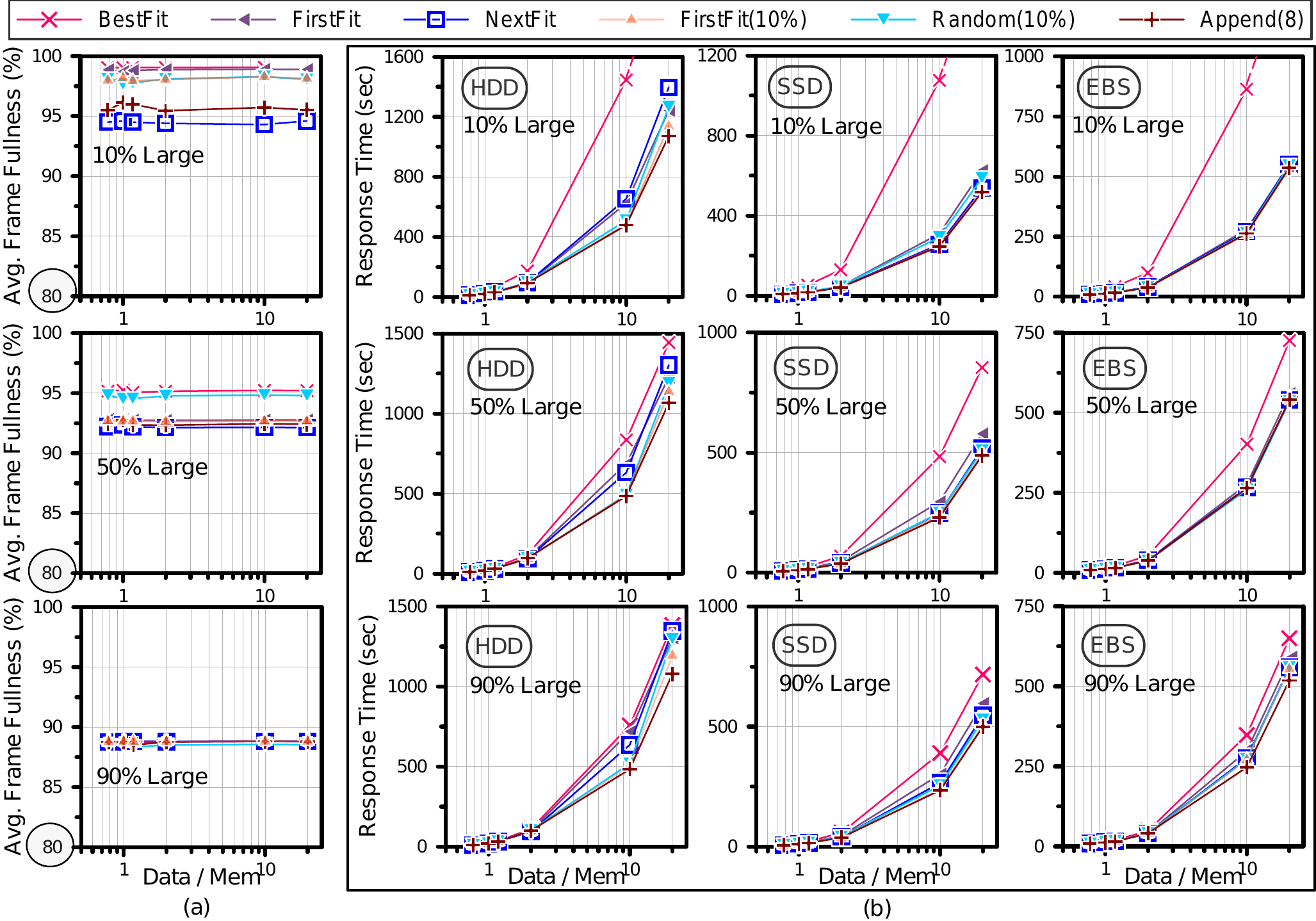}
  \caption{Partition Insertion - 3-Large Record Coexist (a) Average frame fullness (b) Response time on different storage types}
  \label{fig:DI_3L}
  \vspace{-5mm}
\end{figure*}

This section evaluates the performance of different partition insertion algorithms with input datasets containing records of various sizes. 

\textbf{3-Large Coexist.} We use the 3-Large Record Coexist dataset for this experiment. The large records versus small records ratio varies between $10\%$, $50\%$, and $90\%$. As Figure \ref{fig:DI_3L}-(a) shows, increasing the percentage of large records lowers the average frame fullness in all algorithms and minimizes their differences in frame compactness. Inserting large records in a frame may leave a large leftover space that can only be filled with small records. If the small records are limited in number (higher percentage of large records), these leftover spaces remain unfilled and decrease the average fullness. Additionally, the difference between the average frame fullness of the various algorithms diminishes if most of the records are large since only a few frames may have enough space for large records. \newline
As Figure \ref{fig:DI_3L}-(b) shows, Best-Fit again has the highest response time since for each record insertion as it searches all of the in-memory frames of the partition. Best-Fit's response time is worse than the other alternatives with a lower percentage of large records, as more records must be processed. Furthermore, a higher number of records leads to more searching and thus to a higher response time. This rationale is true for the Random algorithm, too, since the random function will be called for $10\%$ of the frames per record insertion. In all of these experiments, Append($8$) has the lowest response time; doing the least amount of work, it still achieves a similar frame fullness to the more intelligent and search-intensive algorithms. While the algorithms other than Append($8$) and Best-Fit perform similarly, the algorithms with a stopping criteria perform slightly better. Storage-wise, the overall response time is higher for HDD than for SSD and Amazon EBS due to its longer time for I/O operations.

\begin{figure*}
  \centering
  \includegraphics[scale=0.80]{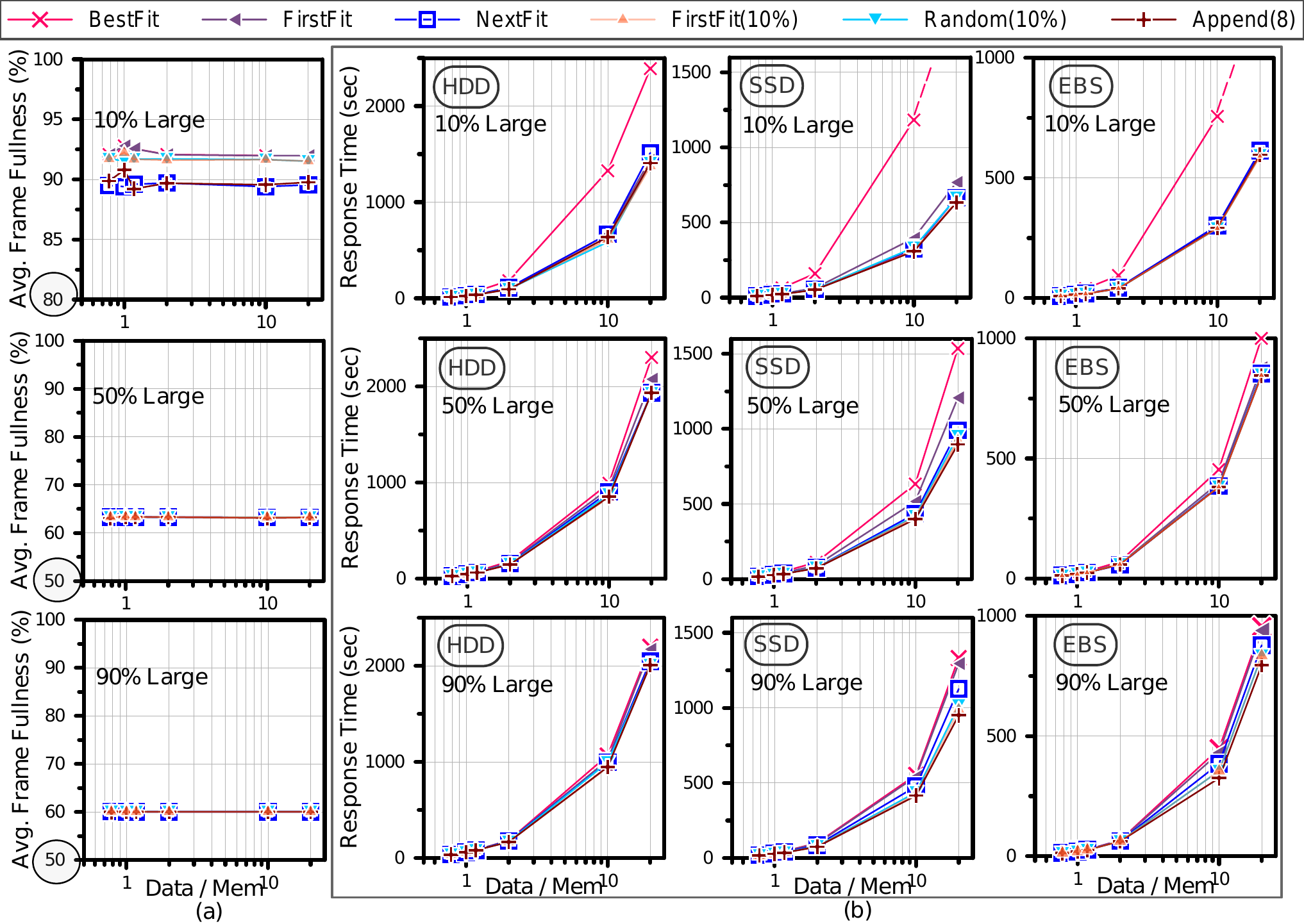}
  \caption{Partition Insertion - 1 Large Record Coexist (a) Average frame fullness (b) Response time on different storage types}
  \label{fig:DI_1L}
\end{figure*}

\textbf{1-Large Coexist.} In the second variation of our experiments for partition insertion with variable-sized records, we use the 1-Large Record Coexist dataset. As above, both inputs are 1GB. Although all of the datasets are 1GB in this and the previous experiment, the datasets for this experiment have a lower cardinality. This is because the large records in this experiment are approximately 3 times larger than the large records in the previous experiment. 
We can observe from Figure \ref{fig:DI_1L}(a) that the frame fullness is higher in cases where most of the records are small, especially in the $10\%$ Large case. This is due to several factors:
\begin{itemize}
    \item The lower percentage of large records means that most of the 1 GB relation is made up of smaller records. Smaller records have a better chance of fitting in partially full frames and thus increasing the frames' compactness.
    \item The higher number of remaining records, especially when they are small, increases the possibility of making the allocated frames more full.
\end{itemize}
We see frame fullness drop from $90\%$ to $62\%$ and $60\%$ as we increase the ratio of large records from $10\%$ to $50\%$ and $90\%$, respectively. This is because each large record requires its own frame, and the small records in the minority can fill the leftover space.\newline
The overall frame fullness in this experiment is lower than in the previous experiment since in this case only one large record can fit in a frame, while in the previous experiment, 3 large records could coexist in one frame and further reduce the leftover space.\newline
As Figure \ref{fig:DI_1L}-(b) shows, similar to the previous experiment, the Best-Fit algorithm has the highest response time and the Append($8$) algorithm has the lowest response time in the majority of the cases. However, the difference between Best-Fit and the other algorithms is not as high as in the 3 Large Record Coexist experiment due to the lower cardinality of the inputs reducing the total search costs.

\section{Growth Policies for Spilled Partitions}\label{GSNGNS}
In the case of insufficient memory, some of the partitions must be written to disk to open up space for additional incoming records. We will consider several victim selection policies - policies which select a memory-resident partition to spill - under two variations of how the memory allocation to spilled partitions is managed:
\begin{enumerate}

\item\textbf{No Grow-No Steal (NG-NS):} There are two main rules for this policy:
\begin{itemize}[leftmargin=*]
\item \textbf{No Grow}: A spilled partition can only have one frame to be used as its output buffer once it has spilled.
\item \textbf{No Steal}: Only memory-resident (unspilled) partitions are selected as victims in case of insufficient memory. A spilled partition writes its output buffer to disk only if the next record hashed to that partition requires more space.
\end{itemize}

\item\textbf{Grow-Steal (G-S):} This growth policy consists of two main rules as well:
\begin{itemize}[leftmargin=*]
    \item \textbf{Grow}: Spilled partitions may grow as large as the available memory lets them.
    \item \textbf{Steal}: Spilled partitions have a higher priority to be chosen as a victim partition in cases of insufficient memory.
\end{itemize}
\end{enumerate}
We compare these growth policies from both analytical and experimental standpoint in the next subsections.
\subsection{Analytical I/O comparison between NG-NS and G-S}
In this section, we look at the I/O differences between the two growth policies for spilled partitions from an analytical point of view. It is important to realize that both policies perform almost the same amount of I/O; however, they differ from one another in their use of random versus sequential I/O. All of the notations used in this section can be found in Table \ref{table:notions}.
\begin{table}[!h]
\begin{center}
\begin{tabular}{|l|l|l|l|l|l| } 
\hline
\thead{Notation} & \thead{Definition} & \thead{Example} \\
\hline
R & Size of build relation in frames & 100 \\
M & Size of memory in frames & 50\\
P & Number of partitions & 20\\
x & Number of spilled partitions & 5\\
\hline
\end{tabular}
\caption{ Notation used in cost formulas}
\label{table:notions}
\end{center}
\vspace{-8mm}
\end{table}

\noindent
\textbf{I/O Analysis for NG-NS.} Let us assume that records are similar in size and that there is no skew in join attribute values. Using this assumption, all partitions are similar in size, in the number of frames, and in the number of records. The following equation calculates the total number of partitions remaining in memory at the end of the build phase:
\begin{equation}\label{inmemPartitions_eq}
P-x = MAX\left(P,\floor*{\frac{M}{\frac{R}{P}}}\right)
\end{equation}
In NG-NS, a memory-resident partition is selected for spilling to disk only when
\begin{enumerate*}
    \item the partitions spilled so far each have a maximum of one frame and,
    \item the in-memory partitions ($P-x$ partitions) have used the rest of the frames ($M-x$) and,
    \item the next incoming record is hashed into a memory-resident partition.
\end{enumerate*}

For our calculation, we can choose any of the partitions to spill as they are all in a similar situation due to the uniformity of data.
By spilling the selected partition, $\frac{M-x}{P-x}$ of this partition’s data is written to disk sequentially, while the rest of its data ($\frac{R}{P} - \frac{M-x}{P-x}$) will later be written to disk randomly (\textit{i.e.,} one frame at a time).\newline
The following equation calculates the amount of temporary results (build phase only) written to disk in a random and sequential fashion under the NG-NS growth policy:\newline
\begin{equation}\label{NGNS_IOinBuild_eq}
\sum_{i=1}^{x}\left(\frac{R}{P} - \frac{M - i + 1}{P - i + 1}\mbox{ Random I/O }\right)+\left(\frac{M - i + 1}{P - i + 1} \mbox{ Seq. I/O }\right)
\end{equation}
\newline
\textbf{I/O Analysis for G-S.} Similar to NG-NS, the next memory-resident partition will spill to disk only if 
\begin{enumerate*}
    \item the incoming record is hashed into a memory-resident partition, and
    \item each spilled partition has at most 1 frame, and
    \item the rest of the memory frames have already been assigned to memory-resident partitions.
\end{enumerate*}

Like NG-NS, each spilling partition writes $\frac{M-x}{P-x}$ of its data frames to disk sequentially when it spills for the first time; however, in contrast to NG-NS, G-S writes the rest of the partition’s frames in chunks consisting of more than one frame.\newline
The second part of Equation \ref{NGNS_IOinBuild_eq} holds for G-S as well, as the next victim partition has $\frac{M-x}{P-x}$ frames in memory. The following equation calculates the sizes of the data chunks written to disk by spilled partitions between the $x\textsuperscript{th}$ and $(x+1)\textsuperscript{st}$ time that a memory-resident partition was selected as a victim.

\begin{equation}\label{GS-chunks}
\frac{1}{P} * \frac{M-x+1}{P-x+1} + \left(\frac{1}{P}\right)^2 * \frac{M-x+1}{P-x+1} + \left(\frac{1}{P}\right)^3 * \frac{M-x+1}{P-x+1} + ...+1
\end{equation}

which reduces to:
\begin{equation}\label{GS-chunks-reduction}
\lim_{P\to\infty} \frac{1}{1-\frac{1}{P}}\left(\frac{M-x+1}{P-x+1}\right)
\end{equation}

Therefore the cost formula in number of I/Os for G-S is:\newline
\small \begin{equation}\label{GS_eq}
 \sum_{i=1}^{x} \left(\lim_{P\to\infty}\frac{1}{1-\frac{1}{P}}\left(\frac{M-i+1}{P-i+1}\right) \mbox{ Seq. I/O } \right) + \left(\frac{M - i + 1}{P - i + 1} \mbox{ Seq. I/O } \right)
\end{equation}
\normalsize
\noindent The first term in Equation \ref{GS_eq} shows that in G-S, each spilled partition writes the rest of its data to disk sequentially. This I/O behavior is different from NG-NS (Equation \ref{NGNS_IOinBuild_eq}), in which the rest of a partition's data is written to disk frame by frame once it first spills.

\subsection{Experimental Analysis of Growth Policies}
\begin{figure*}
  \centering
  \includegraphics[scale=0.70]{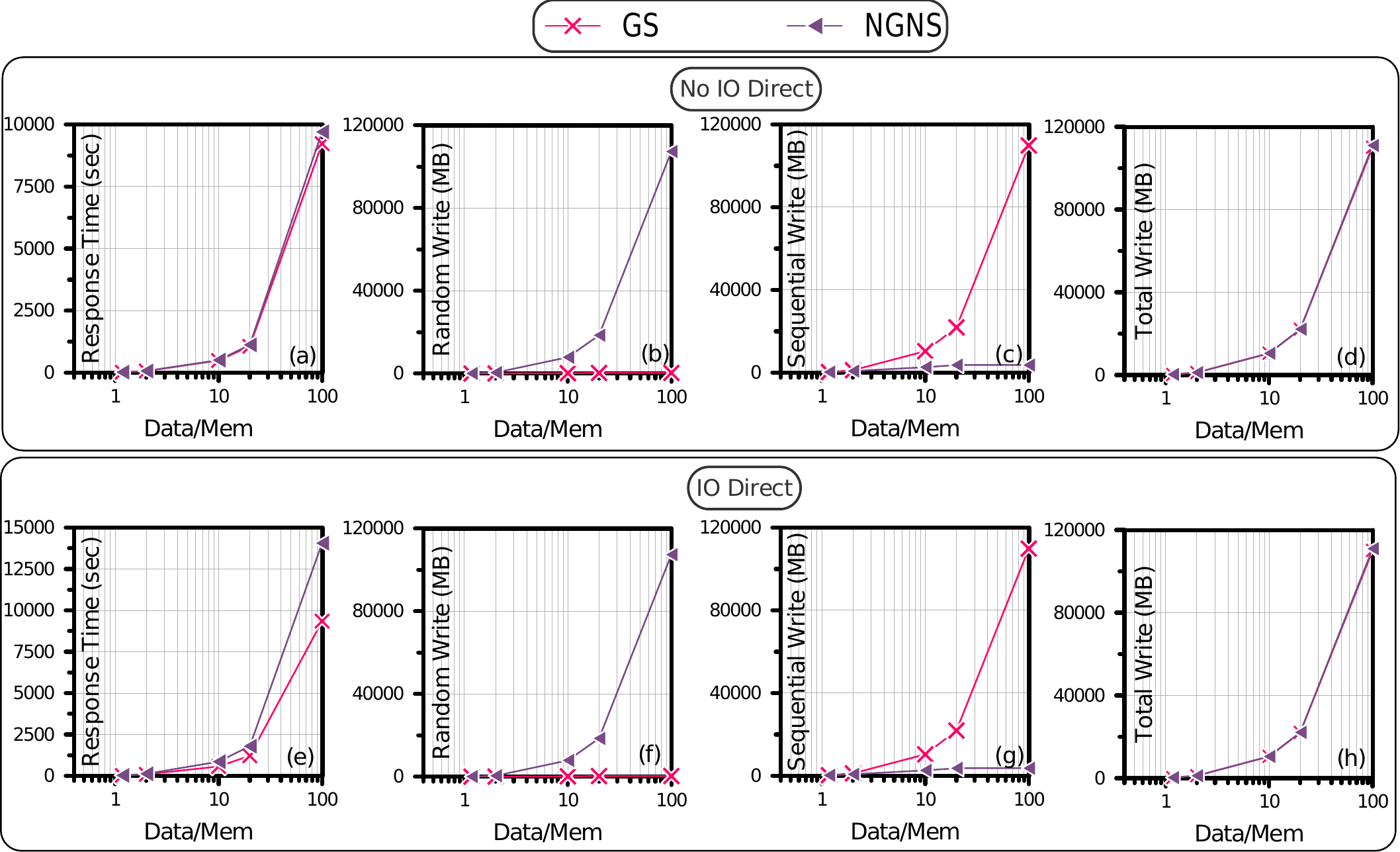}
  \caption{Spilled Partition Growth Policies. (a,b,c,d) - Statistics of GS and NG-NS policies with filesystem cache in use. (e,f,g,h) -  Statistics of GS and NG-NS policies with filesystem cache disabled. }
  \label{fig:GS-NGNS}
  \vspace{-4mm}
\end{figure*}
Based on the cost functions we developed in the previous subsection, we showed that the NG-NS policy leads to more random writes due to using one output buffer allocation per spilled partition. On the other hand, G-S allows the spilled partitions to acquire more than one frame, so its I/O pattern becomes more sequential. Turning random writes into sequential ones can improve performance, especially in systems utilizing HDD. This section compares these two algorithms empirically to verify our expectations from the cost analysis. We used a single join query for which the build and probe datasets contain identical data generated based on the All Small Record dataset configuration. In this experiment, the available memory for the join is a fixed value of 1024MB, while the size of the build and probe inputs varies from 1.2GB, 2GB, 10GB, 20GB, to 100GB. A hard disk is used as the storage device in this experiment.
 
This experiment compares the two growth policies for spilled partitions under two variations of writing to disk: direct or through the filesystem cache. Some database management systems disable the filesystem cache and manage the buffer cache memory themselves instead. We use the IO\_DIRECT Java library \cite{jaydio} for directly writing data to disk and bypassing the filesystem cache in Linux systems. First, we compare G-S and NG-NS based on their volume of writes and their I/O pattern during the build phase. Figures \ref{fig:GS-NGNS}-d and \ref{fig:GS-NGNS}-h show that G-S and NG-NS do the same amount of writing regardless of using or bypassing the filesystem cache. However, as Figures \ref{fig:GS-NGNS}-c and \ref{fig:GS-NGNS}-g show, G-S does up to 120x more sequential writes than NG-NS, while NG-NS does up to 120x more random writes than G-S (Figures \ref{fig:GS-NGNS}-e and \ref{fig:GS-NGNS}-f). This difference in the I/O patterns of the G-S and NG-NS policies while writing the same amount of data to disk aligns with our  results from the previous subsection. Increasing the input sizes causes more spilling to disk, making the difference between these two policies even more significant.
 
Next, we empirically study the performance of these growth policies with and without filesystem cache being present. Figure \ref{fig:GS-NGNS}-e shows the response time of G-S and NG-NS policies when data is written directly to disk (disabled filesystem cache). In this case, NG-NS takes a longer time than G-S to finish due to performing more random writes. The impact of random writes of NG-NS on its performance becomes more significant as the size of the data relative to memory increases; this is because more data is written randomly and the storage device is an HDD. However, Figure \ref{fig:GS-NGNS}-a shows that using a filesystem cache minimizes the difference in response times of these two policies. This is because the filesystem cache collects write requests and orders them based on their target file location on disk (Elevator Algorithm) before sending them to disk; as a result, many of the random writes turn into sequential ones in NG-NS. 
 
Based on our results, choosing the preferred growth policy depends on whether the DBMS performs its own caching or uses the filesystem cache. In AsterixDB, filesystem caching is allowed, so we chose to use NG-NS. As one can imagine, NG-NS may not utilize its given memory as fully as G-S does. However, other operators of the same or other queries could use this leftover memory under a more global memory management policy (which we intend to investigate in our future work).
\section{Victim Selection Policies}\label{VSPolicies}
One or more memory-resident partitions must be written to disk to regain enough space for the incoming records if the available memory is insufficient. In-memory partitions may have different sizes if records have variable sizes or if their distribution between partitions is unbalanced due to skew in join attribute values. In the case of variable-sized partitions, we must decide which partition(s) should spill to disk, considering that we do not know how much data is left to be processed. The partition selected for spilling is called a victim partition, and the policy based on which victim partitions are selected is called the victim selection policy.

In the original HHJ algorithm \cite{DBLP:journals/tods/Shapiro86,10.1145/602259.602261}, one partition is selected upfront (before the query execution) as the in-memory partition, while the rest of the partitions are disk partitions. To ensure that the chosen partition can indeed remain in memory, we must know the sizes of the inputs and the distribution of join attribute values.

As mentioned earlier, the authors of \cite{DBLP:conf/vldb/NakayamaK88} and \cite{DBLP:conf/vldb/GraefeBC98} instead use dynamic destaging to choose victim partitions at runtime. They always select the largest memory-resident partition as the victim partition and limit the spilled partitions to acquiring a maximum of one frame, following the NG-NS growth policy. Neither of these studies considers other victim selection policies or spilled-partition growth policies. Additionally, they do not provide any experiments to show the superiority of their approach.

In the following, we consider 13 possible policies for selecting the next victim partition among non-spilled partitions. These victim selection policies are designed for the NG-NS growth policy. We compare the performance of the 13 policies under different scenarios. Input dataset sizes are unknown to the DBMS during these experiments. The following list describes the victim selection policies that we consider:\newline
\textbf{Largest Size:} Choose the partition with the largest size in memory as a victim to maximize sequential writes and to defer the next spill(s) as long as possible. \newline
\textbf{Largest Records:} Choose the partition with the maximum number of records to spill.\newline
\textbf{Largest Size Self Victim:} Choose the partition into which the record is hashed if it has at least one frame. Otherwise, choose the largest partition to spill.\newline
\textbf{Median Size:} Choose the partition with the median size among all of the memory-resident partitions as the victim partition.\newline
\textbf{Median Records:} Choose the partition with the median number of records to spill.\newline
\textbf{Smallest Size:} Choose the smallest partition with at least one memory frame as the victim partition to avoid overspilling.\newline
\textbf{Smallest Records:} Choose the memory-resident partition with the minimum number of records (>=1) for spilling.\newline
\textbf{Smallest Size Self Victim:} Choose the partition into which the record is hashed to spill if it has any frames. Otherwise, the smallest-size partition will be selected as the victim.\newline
\textbf{Random:} Choose randomly any of the memory-resident partitions as the victim partition.\newline
\textbf{Half Empty:} This victim selection policy starts optimistically by guessing that the remainder of the build input is small and spills the smallest partition. However, it acts pessimistically and spills the largest partition if more than half of the partitions have spilled.\newline
\textbf{Least Fragmentation:} Choose those partitions that have the least amount of fragmentation in their frames, thus trying to reduce I/O.\newline
\textbf{Low High:} Alternate between spilling the smallest and the largest partition. \newline
\textbf{Record Size Ratio:} Choose a partition that holds the smallest number of records among partitions whose size is equal to or exceeds $80\%$ of the largest partition size (low ratio of the number of records to the partition size); this expedites record processing by storing more records in the memory.

\subsection{ Victim Selection Policy Experiments}

In this section, we study the impact of join attribute value skew, variable sized records, and their combination on the different victim selection policies.

\subsubsection{\textbf{Impact of Join skew}}\label{skew}

\begin{figure}[!h]
       \centering
       \includegraphics[scale=0.75]{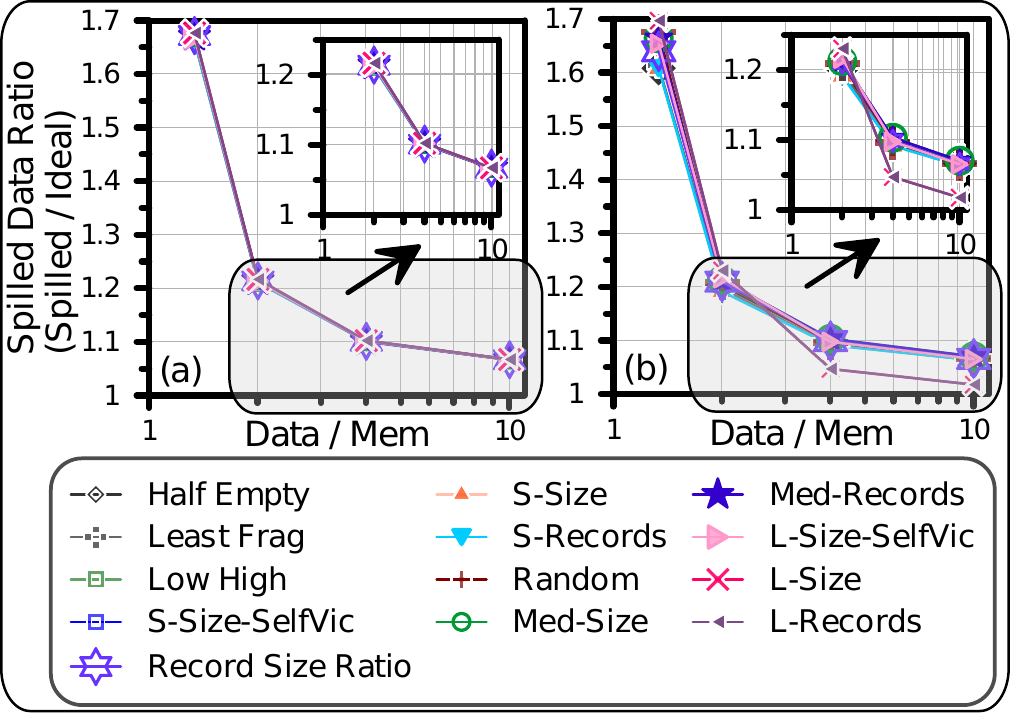}
       \caption{Impact of Join Attribute Value Skew in Victim Selection Policies. (a) - No skew. (b) - Skewed. }
       \label{fig:VS_SKEWNOSKEW}
       \vspace{0mm}
\end{figure}

In our first experiment, we study the impact of join attribute value skew on the 13 different victim selection algorithms. In Figure \ref{fig:VS_SKEWNOSKEW}-a, both the build and probe datasets use the All Small Record configuration, and the join attribute values are unique integers (Non Skewed join attribute value case).

In Figure \ref{fig:VS_SKEWNOSKEW}-b, the join attribute values of the build dataset are integers drawn from a Normal Distribution to make them skewed, while the probe dataset uses unique integers as its join attribute (Skewed join attribute value case). Both relations are 1GB in size and contain $985,000$ records. The authors of \cite{DBLP:conf/sigmod/SchneiderD89, DBLP:conf/tpctc/BonczAK17} used a Normal Distribution in which $99\%$ of the join attribute values are coming from $5\%$ of the possible values, justifying this as similar to the skew found in real-world data. To achieve this data skew, we use a Normal Distribution on an integer attribute with the mean of $492500$ (equal to half of the cardinality), a standard deviation of $8208$, and a range of possible values varying from 1 to the dataset cardinality.
\begin{figure*}
        \centering
        \includegraphics[scale=0.75]{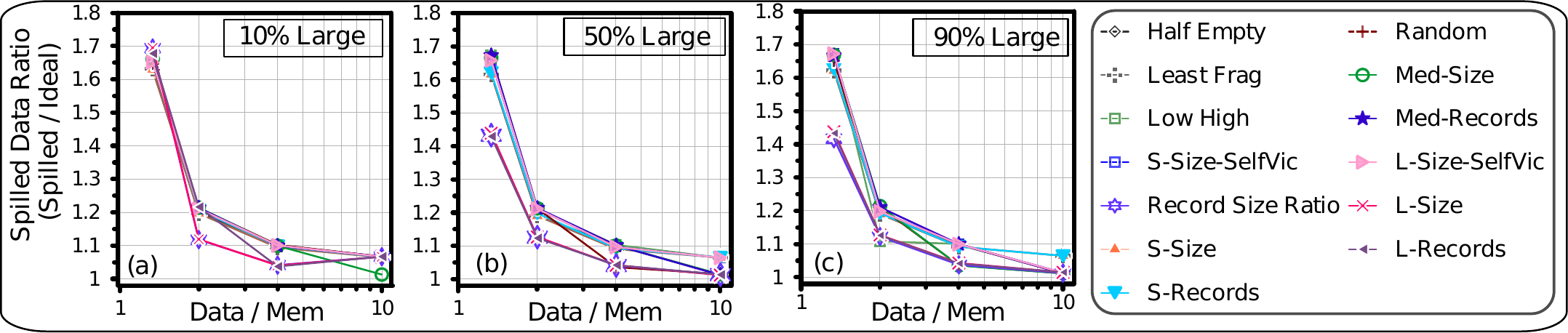}
        \caption{Impact of Variable Record Size (1-Large Record Coexist) in Victim Selection Policies. (a,b,c) - Spilled Data Ratio when $10\%$, $50\%$, and $90\%$ of the records are large, respectively.}
        \label{fig:VS_1L}
 
\end{figure*}

\begin{figure*}
        \centering
        \includegraphics[scale=0.75]{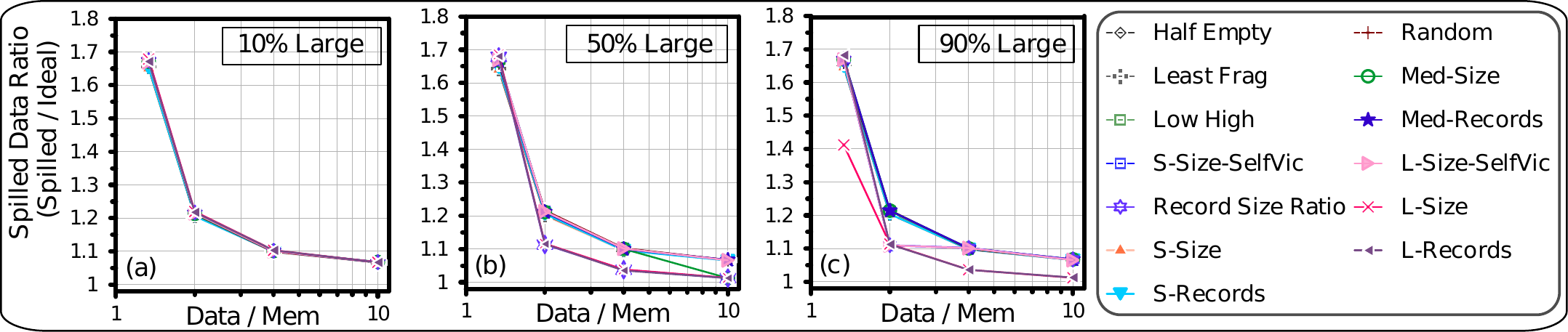}
        \caption{Impact of Variable Record Size (3-Large Records Coexist) in Victim Selection Policies. (a,b,c) - Spilled Data Ratio when $10\%$, $50\%$, and $90\%$ of the records are large, respectively.}
        \label{fig:VS_3L}
        \vspace{0mm}
\end{figure*}
The metric used in Figure \ref{fig:VS_SKEWNOSKEW} is the ratio of the amount of spilled data over the ideal amount of spilling. The ideal amount of spilling is the minimum amount of data that must be spilled to disk during the build phase. We determine the ideal amount of spilling by using a simple simulator program. This simulator minimizes the data spilling by maximizing the memory usage in each round of HHJ with an in-memory partition, similar to the original HHJ operator provided with an accurate a priori information. This simulator a fudge factor of 1.4 to consider the possible fragmentation in frames.

As Figure \ref{fig:VS_SKEWNOSKEW}-a shows, all of the algorithms have a similar performance if records are similar in size and the join attribute values are uniformly distributed. Figure \ref{fig:VS_SKEWNOSKEW}-b shows that skew in the join attribute values can cause different spilling behavior for some victim selection policies.

In Figure \ref{fig:VS_SKEWNOSKEW}-b, the Largest-Size and Largest-Record policies overspill when data is slightly larger than the available memory. However, as the ratio of data over memory increases, spilling the larger partitions releases more frames, saving other partitions from spilling. 

The Smallest-Size and Smallest-Record policies, which spill less data initially, will spill more when the ratio of data to memory is higher. All other policies show a spilling behavior that lies between these two categories of policies. However, the overall difference between most of the policies is almost insignificant.

\subsubsection{\textbf{Impact of Variable-Sized Records}}

Next, we study the impact of variable-sized records on the performance of the victim selection policies. We used a set of 1GB relations based on the 1-Large Record Coexist and 3-Large Record Coexist dataset configurations.
As Figures \ref{fig:VS_1L} and \ref{fig:VS_3L} show, most of the policies perform similarly as the ratio of data over memory is increased in both experiments. The Largest-Size and Largest-Record policies spill less data and fewer partitions to disk than the other victim selection policies in most of the data points. This is because the number of frames that larger partitions free can save more partitions from spilling.

 In both Figures \ref{fig:VS_1L} and \ref{fig:VS_3L}, increasing the population of large records leads to a larger difference between victim selection policies. The variations in the size of the records and the high impact of large records on the partitions' sizes make it possible to see differences between these victim selection policies. In both 1-Large Record Coexist and 3-Large Record Coexist cases (Figures \ref{fig:VS_1L} and \ref{fig:VS_3L}), the Largest-Size, Largest-Records, and in some cases Largest-Size-Smallest-Record policies spill the least amount of data and the fewest number of partitions in most of the data points by spilling the largest partitions first. 
This difference between policies in the 3-Large Record Coexist experiment is less obvious since the large records are $1/3$ of the size of the large records in 1-Large Record Coexist dataset. In Figure \ref{fig:VS_1L}-a most policies perform similarly as there are fewer large records thus, less opportunity for these policies to perform differently.

\subsubsection{\textbf{Impact of Join Skew \& Variable-Sized Records}}
\begin{figure*}
         \centering
         \includegraphics[scale=0.75]{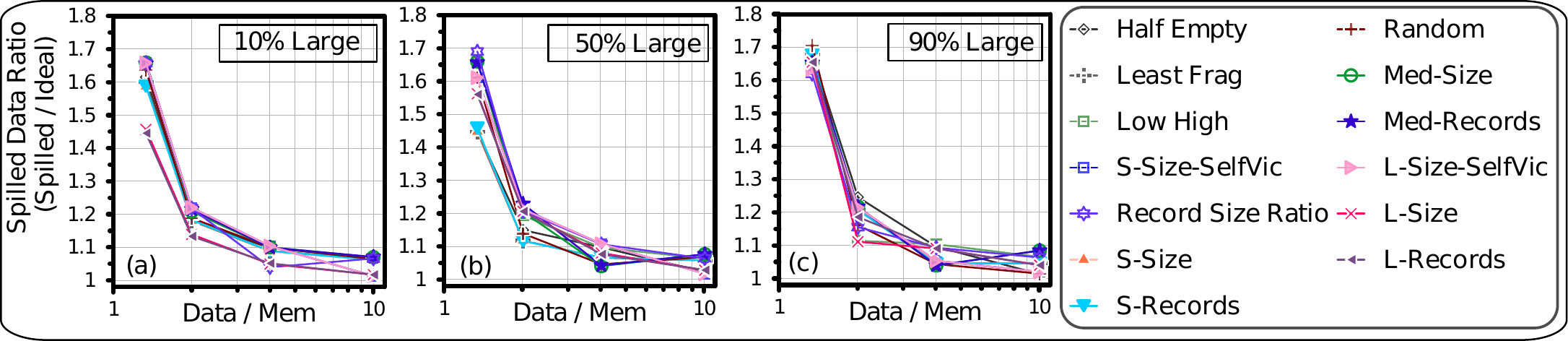}
         \caption{Impact of Skew \& Variable Record Sizes (1-Large Record Coexist) in Victim Selection Policies. (a,b,c) - Spilled Data Ratio when $10\%$, $50\%$, and $90\%$ of the records are large, respectively.}
         \label{fig:VS_1L_Skew}
         \vspace{0mm}
    \end{figure*}

\begin{figure*}
         \centering
         \includegraphics[scale=0.75]{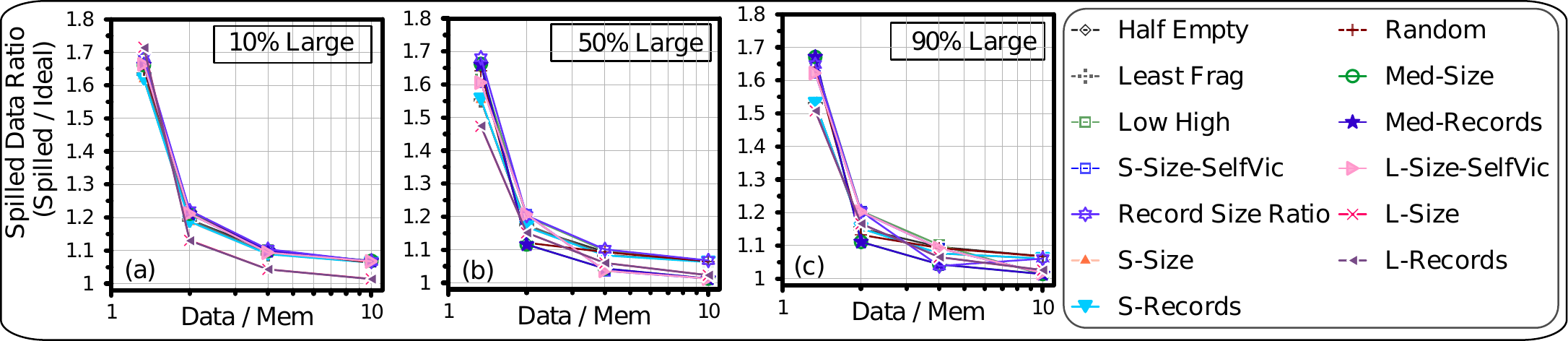}
         \caption{Impact of Skew \& Variable Record Sizes (3-Large Records Coexist) in Victim Selection Policies. (a,b,c) - Spilled Data Ratio when $10\%$, $50\%$, and $90\%$ of the records are large, respectively.}
         \label{fig:VS_3L_Skew}
         \vspace{0mm}
    \end{figure*}

In this experiment, we study the impact of the combination of join attribute value skew and variable-sized records on the different victim selection policies. The same Normal distribution discussed in Section \ref{skew} is used for making the build dataset skewed, and the record sizes are chosen from the same distribution used for 1-Large Record Coexist (Figure \ref{fig:VS_1L_Skew}) and 3-Large Record Coexist (Figure \ref{fig:VS_3L_Skew}) cases. The probe inputs have the same cardinality and record size distribution as the build input, while their join attributes are unique integers. No correlation exists between the record sizes and the join attribute value distribution.
    
Similar to the previous experiment, Largest-Size and Largest-Record are two well-performing policies when larger records have a lower population. In contrast, the Median Size and Median Records policies perform well by taking a middle route if data is skewed and most of the records are large. The skew in data makes some partitions get more records; partitions with more records will have larger sizes if records are mostly large-sized, and thus the Largest-Size and Largest-Record algorithms can overspill. In the case of very limited memory for the 1-Large Record Coexist case (the first data point in Figure \ref{fig:VS_1L_Skew}-a, \ref{fig:VS_1L_Skew}-b, and \ref{fig:VS_1L_Skew}-c), Smallest-Records and Smallest-Size are two of the best performing policies next to Largest-Size and Largest-Records. Since most of the data is located in a few partitions, there are many small partitions with only a few frames. As such, Smallest-Records and Smallest-Size can avoid overspilling by spilling these small partitions when data is just slightly larger than memory.

In the 3-Large Records Coexist case, the victim selection policies' performance is similar to the 1-Large Record Coexist case with the difference that algorithms such as Median Records also perform well in this case due to the smaller sizes of large records. As both Figures \ref{fig:VS_1L_Skew} and \ref{fig:VS_3L_Skew} show, the victim selection policies perform similarly in terms of how much data they spill to disk; however, they differ in their I/O patterns.
Algorithms such as Largest-Size and Largest-Records, tend to write larger numbers of frames sequentially; others such as Smallest-Size and Smallest-Records write a smaller number of frames in a more random manner. As our experiments for G-S and NG-NS showed, this difference in their I/O patterns may not impact performance as much as otherwise expected if filesystem caching is enabled. 

\subsection{Results for Victim Selection Policy}
Based on our experiments in the previous subsection, in most cases, especially when data is much larger than memory (the Big Data world), the Largest-Size and Largest-Record policies do less I/O than the other victim selection policies. Our results confirm the conjecture of \cite{DBLP:conf/vldb/NakayamaK88,DBLP:conf/vldb/GraefeBC98} that the Largest-Size policy (as well as the Largest-Record policy, based on our results) is a good selection policy for the following two reasons:
\begin{enumerate*}
    \item Larger partitions release many frames; thus, they save other partitions from spilling to disk. This leads to less data being written to disk.
    \item Writing larger partitions leads to more sequential and less random writes. 
\end{enumerate*}

However, our results also show that the difference in the amount of spilled data is not enough to make a significant difference in performance. Although spilling larger partitions leads to a more sequential I/O pattern, its benefits in performance are diminished if filesystem caching is enabled.

\section{Other Optimization Techniques}\label{Optimization}
There are a few other techniques that AsterixDB uses in order to improve the performance of Hybrid Hash Join which we discuss in this section.
\subsection{Switch to Block Nested Loop Join}
If the skew in the join attribute values is too high, hashing can become ineffective and as such, regardless of the number of rounds that the Hybrid Hash Join operator executes, it may not finish in time, or not at all in some cases. As such, in AsterixDB, in each round of Hybrid Hash Join after the initial round, the size of the build input (smaller input) will be compared to its size in the previous round. If the reduction in size was less than $20\%$, the data is inferred to be highly-skewed, and a block nested loop join will subsequently be used instead of Hybrid Hash Join. In \cite{DBLP:conf/vldb/GraefeBC98} this technique is called \textit{Bail-out}.

\subsection{Role Reversal of Input Relations}
In each round of Hybrid Hash Join after the initial round, the size of each input partition is known since it was read into memory and spilled in the previous round. As such, Hybrid Hash Join can now use this information and choose the smaller input to its operator to participate in the build phase and the other relation in the probe phase. This selection can make the build relation in the previous round become the probe relation in the following round, which is called role reversal. This technique can help with reducing the total volume of reads and writes during the execution of the join operator by choosing the smaller relation as the build relation.

\subsection{In-memory Hash Join}
After the first round of Hybrid Hash Join, the sizes of the inputs of the next rounds of hybrid hash join are known as they were written to the disk in the previous round of hybrid hash join. If the build relation (or the smaller input in case of role reversal) and its hash table can fit inside the memory, the hybrid hash join algorithm skips the partitioning phase altogether and directly creates the hash table and fills it by processing the records of the build relation. In this way, we avoid the unnecessary task of partitioning and we save CPU cycles which can reduce response time, especially if the underlying storage hardware is SSD (where CPU cost becomes more prominent).

\subsection{Best Match}
As after the first round of the Hybrid Hash Join operator, the size of the input relations to the join is known, so the join can use this information for selecting victim partitions more informatively. For example, it can choose the partition with the closest size to the remainder of the build relation to spill in order to avoid overspilling.

\subsection{Reloading Spilled Partitions}
We can estimate the size of the hash table at the end of the build phase based on the cardinality of the in-memory partitions. Based on this estimation and before creating the hash table, we try to bring back those spilled partitions that can fit in the leftover join memory. In this way, we can reduce the amount of I/O during the probe phase by bringing the corresponding partitions from the build phase back to memory.

\section{Conclusion}\label{conclusion}
Our experimental study has investigated different policies and optimization techniques to design a robust Dynamic HHJ operator when no a priori information about the input datasets and their value distributions is available. 

One of the steps in configuring the HHJ operator is to determine the number of the partitions that the build and probe inputs should be hashed into. While previous studies have suggested an upper bound for the number of partitions, no lower bound or default value for the number of partitions has been proposed to the best of our knowledge. Not having a reasonable lower bound can lead to having too few partitions, causing overspilling. Based on a simulation study, we recommended 20 as the minimum number of partitions and also as the default number of partitions when no information on the input sizes exists. With at least 20 partitions, each spilled partition writes only $5\%$ or less of the build input to disk if the join attribute values are uniformly distributed. 

Furthermore, we explored different partition insertion algorithms for incoming records to find a frame with enough space among a partition’s in-memory frames. It is important to find frames with enough space in a partition efficiently while still making the frames as full as possible. Append($8$) showed the best performance among the partition insertion algorithms. Its partition fullness was only slightly lower than the more guided and extensively searching algorithms. However, Append($8$) had a better response time and achieved a similar partition fullness with fewer frames checked.

Next, we considered two potential post-spilling growth policies for spilled partitions, Grow-Steal and No Grow-No Steal. Our cost model showed that Grow-Steal should perform better than No Grow-No Steal due to doing more sequential I/Os. Our experiments showed that this behavior indeed happens when the file system cache is not in the picture, but we saw that a modest file system cache can mitigate this difference by turning most random I/Os into sequential ones.

Additionally, we considered 13 different partition victim selection policies and evaluated their performance under different scenarios of record sizes and join attribute value skew. Our results confirmed the conjecture in previous work that the Largest Size policy is one of the best policies in most cases. However, this difference is not large enough to impact the performance of the overall system. Although victim selection policies differ with each other in their I/O pattern, this difference will be diminished if filesystem caching is enabled. 

\section*{Acknowledgement}
This work has been supported by NSF awards IIS-1838248, CNS-1925610, and IIS-1954962 along with industrial support from Google and support from the Donald Bren Foundation (via a Bren Chair).


\bibliographystyle{ACM-Reference-Format}
\bibliography{sample}

\end{document}